\documentclass[a4paper,11pt]{article}

\usepackage{jheppub} 

\usepackage[T1]{fontenc} 
\usepackage{amssymb}
\usepackage{amsmath}
\usepackage{nicefrac}
\usepackage{nccmath}
\usepackage{amsfonts}                   
\usepackage{mathtools}
\usepackage{bbold}
\usepackage{marvosym}
\usepackage{physics}
\usepackage{listings}
\usepackage{relsize}
\usepackage{gauss}
\usepackage{empheq}

\usepackage{enumitem}
\usepackage{pdfpages}
\usepackage{graphics}
\usepackage{graphicx}
\usepackage{color}  
\usepackage{soul}  
\usepackage{rotating}
\usepackage{svg}
\usepackage{tikz}
\usepackage{xcolor}
\usepackage{pstricks}
\usepackage{transparent}

\usepackage[english]{babel}
\usepackage{scrextend}
\usepackage{multirow}
\usepackage[draft]{todonotes}
\usepackage{dirtytalk}
\usepackage[style=english]{csquotes}
\usepackage{tocloft}
\usepackage{ocg-p}
\usepackage{ocgx}
\usepackage{floatrow}
\usepackage{sidecap}
\usepackage{comment}
\usepackage[numbers]{natbib}
\usepackage{epigraph}

\usepackage[bottom]{footmisc}
\usepackage{slashed}
\usepackage{multicol}
\usepackage{fancyhdr}
\usepackage{verbatim}
\newcounter{eg}                         \newtheorem{eg}{Example}[section]        
\def\beg{\begin{eg}\rm}                 \def\eeg{\hfill\sq\end{eg}}

\DeclareMathOperator{\0}{\mathbb{0}}
\DeclareMathOperator{\1}{\mathbb{1}}
\DeclareMathOperator{\diag}{diag}
\DeclareMathOperator{\col}{col}
\DeclareMathOperator{\row}{row}
\DeclareMathOperator{\sgn}{sgn}
\DeclareMathOperator{\func}{func}

\newcommand{\Var}{\operatorname{Var}} 
\newcommand{\obs}{\mathcal{O}}

\newcommand{\wt}{\widetilde}

\def\lr#1{\left(#1\right)}
\def\avg#1{\left\langle #1 \right\rangle}
\def\K{\mathcal K}

\usepackage{url}
\usepackage{xurl}
\PassOptionsToPackage{hyphens}{url}
\usepackage{hyperref}
\hypersetup{
    colorlinks=true,
    citecolor=[rgb]{0,0.334,0.666}, 
    linkcolor=[rgb]{0,0.334,0.666},
    urlcolor=[rgb]{0,0.334,0.666},
}

\title{\boldmath Approximate treatment of noncommutative curvature in quartic matrix model}

\author[a,b]{D. Prekrat,
}
\author[b]{D. Rankovi\'{c},}
\author[b]{N. K. Todorovi\'{c}-Vasovi\'{c},}
\author[c,d]{S. Kov\'{a}\v{c}ik,}
\author[c]{and J. Tekel}

\affiliation[a]{University of Belgrade -- Faculty of Physics,\\ P.O. Box 44, 11001 Belgrade, Serbia}
\affiliation[b]{University of Belgrade -- Faculty of Pharmacy,\\ Vojvode Stepe 450, Belgrade, Serbia}
\affiliation[c]{Department of Theoretical Physics, Faculty of Mathematics, Physics and Informatics, \\
Comenius University in Bratislava, \\
Mlynsk\'a dolina, 842 48, Bratislava, Slovakia}
\affiliation[d]{Department of Theoretical Physics and Astrophysics, Faculty of Science, Masaryk University, \\
Brno, Czech Republic}

\emailAdd{dprekrat@ipb.ac.rs}

\abstract{
We study a Hermitian matrix model with the standard quartic potential amended by a $\mathrm{tr}(R\Phi^2)$ term for fixed external matrix $R$. This is motivated by a curvature term in the truncated Heisenberg algebra formulation of the Grosse-Wulkenhaar model---a renormalizable noncommutative field theory. The extra term breaks the unitary symmetry of the action and leads, after perturbative calculation of the unitary integral, to an effective multitrace matrix model. Accompanying the analytical treatment of this multitrace approximation, we also study the model numerically by Monte Carlo simulations. The phase structure of the model is investigated, and a modified phase diagram is identified. We observe a shift of the transition line between the 1-cut and 2-cut phases of the theory that is consistent with the previous numerical simulations and also with the removal of the noncommutative phase in the Grosse-Wulkenhaar model.
}

\begin{document} 
\maketitle
\flushbottom

\section{Introduction}

qMatrix models have broad and diverse application in physics: from biophysics, solid state physics, optics, nuclear physics to quantum gravity \cite{Eynard:2015aea,Akemann:2011csh,PhysRevLett.94.168103,Beenakker:2014zza,Weidenmuller:2008vb,Guhr:1997ve,Loll:2019rdj}. Degrees of freedom of these models are expressed in terms of matrices of a certain type, for example, Hermitian matrices, real symmetric matrices, etc. Matrix models can serve as a tool to regularize the given theory \cite{Bietenholz:2014sza}. It is also conjectured that they express fundamental laws of nature. Also, while in some models the matrix elements are functions on some spacetime, in others, there is no explicit notion of spacetime. Finally, in some intriguing cases, the very notion of spacetime emerges, at least in a portion of the model's parameter space \cite{Steinacker:2011ix}.

From a mathematical perspective, among the best understood is the class of pure potential matrix models in which the action is specified as a (multi)trace of some polynomial of matrices. Perhaps the simplest example is the Gaussian Unitary Ensemble where the degrees of freedom are ${N \times N}$ Hermitian matrices $\Phi$ and the mean values $\expval{\obs}$ of observables $\obs$ are expressed as 
\begin{equation}
    \left\langle \obs\!\left( \Phi \right) \right\rangle = \frac{1}{Z} 
   \int [d \Phi]\, \obs\!
    \lr{\Phi} e^{- B \tr \Phi^2},
    \qquad\qquad
    Z = \int [d \Phi]\, e^{- B \tr \Phi^2}.
\end{equation}

The integration $\int [d \Phi]$ over $N^2$ real parameters of the matrix elements can be split into integration over $N$ eigenvalues and a unitary matrix $U$. This change of variables and integration over $U$ generates the logarithmically-repulsive Vandermonde term
\begin{equation}
  S_\text{\smaller VDM} =  - \sum \limits_{i \neq j} \log\abs{\lambda_i - \lambda_j} \, ,
\end{equation}
which is responsible for most of the important properties of matrix models. If only functions of the eigenvalues $\lambda_i$ of $\Phi$, for example $\tr \Phi^2 = \sum_i \lambda_i^2$, are of interest, the $[dU]$ integral is trivial and the problem reduces to the $N$ dimensional problem of eigenvalues.

There are two different approaches to calculating the expectation values and other quantities in matrix models. The first uses various analytical techniques \cite{Eynard:2015aea,Zuber2012IntroductionTR} (e.g., saddle point approximation, orthogonal polynomials, loop equations) to obtain exact results. These are, however, usually applicable for rather simple probability distributions and are less useful in more complicated models. The second computes the multidimensional integral numerically, usually by using variations of the Monte Carlo method \cite{Wolff:1988uh,Duane:1987de,Ydri:2015zba,handbookMCMC}. The advantage of this approach is the ability to treat basically any probability distribution; the disadvantage is that it provides only uncertain information due to numerical procedures and only for specific numerical values of the distribution parameters and limited matrix sizes.

Usually, the pure potential term is complemented by additional terms providing a physical context. For example, many models include a kinetic term that is related to the geometry of noncommutative (NC) space, and the matrix model then describes a field theory defined on such a space \cite{Balachandran:2005ew,Ydri:2016dmy}. The prime example is the fuzzy sphere, which is a regularization of the ordinary sphere with a finite number of degrees of freedom that is constructed using finite-dimensional representations of the $su(2)$ algebra \cite{Madore:1991bw,hoppe}.

The additional terms, however, make the change of variables less useful–––even if we are only interested in functions of the form $\obs(\Lambda)$, the action now depends not only on the eigenvalues $\lambda_i$ but on all elements of $\Phi$ and the $[dU]$ integration is no longer trivial. Considerable efforts were made to study how the additional terms can be approximated in a way that would allow us to solve this problem. For example, the kinetic term of the fuzzy sphere model can be reasonably well replaced by a simpler multitrace term \cite{OConnor:2007ibg,Subjakova:2020prh}. 

In this paper, we propose a new way of dealing with the terms in the action that make the angular integration difficult by means of the Harish-Chandra-Itzykson-Zuber (HCIZ) integral \cite{Zuber2012IntroductionTR,Eynard:2015aea}. To show the efficacy of this approach, we will investigate a model---obtained from the Grosse-Wulkenhaar (GW) model \cite{Grosse:2003nw,Buric:2009ss,Prekrat:2021uos} by removing the kinetic term---in which the matrix $\Phi$ is coupled to a background matrix $R$ describing the curvature of the underlying NC space. More details about this model and some previous results are given in the next section.

This paper is organized as follows. First, in Section \ref{section:GW}, we provide a necessary insight into the GW matrix model.  Next, in Section \ref{section3}, we discuss the HCIZ integral and use it to approximate the curvature contribution to the GW action. In Section \ref{section4}, we show the analytical treatment of this multitrace model, while in Section \ref{section:numerical}, we show the result of numerical studies of the full starting model with the curvature term. The last section is devoted to conclusions, followed by appendices with some examples and technical details.
\section{Grosse-Wulkenhaar matrix model}
\label{section:GW}

Grosse and Wulkenhaar (GW) solved the (non)renormalizability problem of the NC $\lambda\phi^4_\star$ model by enhancing it with a harmonic-oscillator potential term \cite{Grosse:2003nw, Grosse:2004yu,Vinas:2014exa}, which introduces symmetry between large and small energy scales and resolves the problem with UV/IR mixing. As shown by Buri\'{c} and Wohlgennant \cite{Buric:2009ss}, this special term can be reinterpreted as coupling with the curvature of the background NC space, with the GW model obtained in the large-$N$ limit. We will work here with a matrix version of this reinterpreted action---in two dimensions and with rescaled parameters\footnote{Action $\wt{S}$ in \cite{Prekrat:2021uos}.}---which was previously numerically studied in \cite{Prekrat:2020ptq,Prekrat:2021uos}. For simplicity, we will refer to it as the GW model.

Let us start the model building with the usual pure-potential (PP) contribution to the action\footnote{The explicit factor $N$ in the action ensures that the eigenvalue distribution of the matrix $\Phi$ converges to a finite support. To simplify notation, the kinetic term, the NC coordinates, and the curvature are also rescaled compared to \cite{Prekrat:2021uos}.}
\begin{equation}
S_{\text{\smaller PP}} = N\tr \left( - g_2\Phi^2 + g_4\Phi^4 \right),
\end{equation}
in which the field $\Phi$ is a $N\times N$ Hermitian matrix. This is a quartic interaction potential with a negative mass-term. Usually, this term is accompanied by a kinetic term that encodes some underlying geometrical structure so the model can approximate an interacting field theory on some space, usually in the infinite matrix size limit \cite{Balachandran:2005ew,Ydri:2016dmy}. In the case of the GW model \cite{Buric:2009ss,Prekrat:2020ptq,Prekrat:2021uos}, the kinetic term is defined as
\begin{equation}\label{kinetic}
S_{\text{\smaller K}} = N\tr\Phi\K\Phi \, ,
\qquad\quad
\mathcal{K}\Phi = \comm{X}{\comm{X}{\Phi}} + \comm{Y}{\comm{Y}{\Phi}} \, ,
\end{equation}
where $X$ and $Y$ are rescaled NC coordinates\footnote{Actually, the truncated Heisenberg algebra is 3-dim but gets \enquote{compactified} into the $XY$-plane in the $N\to\infty$ limit, in which we are interested. For more details, see \cite{Buric:2009ss}.} of the truncated Heisenberg algebra 
\\
\begin{equation}
X = \frac{1}{\sqrt{2N}}
\begin{pmatrix} 
 & \scriptstyle+\sqrt{1} \\ 
 \scriptstyle+\sqrt{1} &  & \scriptstyle+\sqrt{2} \\ 
 & \scriptstyle+\sqrt{2} & &  \hspace{-25pt}\begin{rotate}{-5}{$\ddots$}\end{rotate} \\ 
 &  & \hspace{-25pt}\begin{rotate}{-5}{$\ddots$}\end{rotate} &  & \hspace{-16pt}\scriptstyle+\sqrt{N-1} \\ 
 &  & & \hspace{-15pt}\scriptstyle+\sqrt{N-1} &
\end{pmatrix},
\qquad
Y = \frac{i}{\sqrt{2N}}
\begin{pmatrix} 
 & \scriptstyle-\sqrt{1} \\ 
 \scriptstyle+\sqrt{1} &  & \scriptstyle-\sqrt{2} \\ 
 & \scriptstyle+\sqrt{2} & & \hspace{-25pt}\begin{rotate}{-5}{$\ddots$}\end{rotate} \\ 
 &   & \hspace{-25pt}\begin{rotate}{-5}{$\ddots$}\end{rotate} &  & \hspace{-16pt} \scriptstyle-\sqrt{N-1} \\ 
 &  &    & \hspace{-15pt}\scriptstyle+\sqrt{N-1} &
\end{pmatrix}.
\end{equation}
\\
\noindent
As already mentioned, this background space has a curvature, here rescaled to
\begin{equation}
R=\frac{15}{2N}-8\lr{X^{2}+Y^{2}},
\end{equation}
that is, more explicitly,
\begin{equation}
R=\frac{31}{2N} - \frac{16}{N}\diag\lr{1,2,\ldots,N-1,N/2}.
\end{equation}
To make calculations easier and arguments more transparent, we shift $R$ by $-31/(2N)$, which changes the mass parameter $g_2$ by $-31g_r/(2N)$. We also replace $R_{NN}\to -16$, since neither of these changes affects the large-$N$ limit. As a result, we work with
\begin{equation}
    R = -\frac{16}{N} \diag\lr{1,2,\ldots N}.
    \label{curvature}
\end{equation}
The curvature term is then 
\begin{equation}
S_{\text{\smaller R}} = N\tr\lr{- g_r R\Phi^2},
\label{curvature term}
\end{equation}
and finally, the full action of the GW model is given by
\begin{equation}
S_{\text{\smaller K}+\text{\smaller R}+\text{\smaller PP}} = N\tr\lr{
\Phi\mathcal{K}\Phi
- g_r R\Phi^2
- g_2\Phi^2 + g_4\Phi^4
}.
\label{GW matrix model}
\end{equation}
We will consider the large-$N$ limit of this model while keeping the implicit NC length scale fixed.

As shown numerically in \cite{Prekrat:2020ptq,Prekrat:2021uos}, the model \eqref{GW matrix model} hosts three phases which meet at a triple point: standard (a)symmetric 1-cut\footnote{In further text, we will denote the phases as 1-cut-sym, 1-cut-asym and 2-cut-sym.} phases and a NC symmetric 2-cut/striped phase sandwiched in between. This striped phase is believed to be behind the phenomenon of UV/IR mixing. With no curvature term, the triple point of this model lies at the origin of the phase diagram. Once added, the curvature pushes the transition lines toward the larger $g_2$, proportionally to the curvature coupling $g_r$. This ensures that the bare $g_2$ starts in the 1-cut-sym phase, therefore solving the UV/IR problem. We will here attempt to derive this transition-line shift analytically.

Treatment of various forms of the kinetic term has been studied in the literature \cite{OConnor:2007ibg,Samann:2014hjr,Polychronakos:2013nca,Tekel:2017nzf}. In this paper, we focus on a novel treatment of the curvature term---and the features it leads to---so we drop the kinetic term and for the rest of the paper work with the action
\begin{equation}
S = N\tr\left(
- g_r R\Phi^2
- g_2\Phi^2 + g_4\Phi^4
\right).
\label{matrix model}
\end{equation}

Let us first qualitatively assess the possible classical vacua of the above action. The classical EOM for $S$ is given by
\begin{equation}
-g_r \lr{R \Phi + \Phi R}
+ \Phi\lr{-2g_2+4g_4\Phi^2} = 0 \, .
\end{equation}
If we consider only diagonal solutions, this equation simplifies to
\begin{equation}
\Phi\lr{-g_rR-g_2+2g_4\Phi^2} = 0 \, ,
\end{equation}
and is solved by
\begin{equation}
\Phi^2 = \frac{g_2\1+g_r R}{2g_4}
\qquad
\vee
\qquad
\Phi = \0 \, .
\end{equation}
Since $R<0$, the nontrivial solution is defined for
\begin{equation}
    \label{bound1}
    g_2 \geq \max_i\abs{R_{ii}} \cdot g_r=16g_r
\end{equation}
where it has lower classical energy. For $g_2 \gg g_r$, its square roots approach the vacua present in the PP model, indicating a similar phase structure. 

Before we proceed to a detailed treatment, we could replace the curvature with its minimal and maximal eigenvalue as in \cite{Prekrat:2020ptq}, therefore bounding our action from both above and below. We can now, naively, expect that our model's transition line lies between the transition lines for these two PP models with shifted masses, that is between $g_2 = 2\sqrt{g_4}$ and $g_2 = 2\sqrt{g_4} + 16g_r$. This gives us an upper bound on the starting point of the transition line
\begin{equation}
    \label{bound2}
    g_2\bigg|_{g_4=0} \leq 16g_r \, .
\end{equation}
Combining \eqref{bound1} and \eqref{bound2} fixes its position to
\begin{equation}\label{16gr}
    g_2\bigg|_{g_4=0} = 16g_r \, .
\end{equation}
As we will see, this agrees with the results of numerical simulations but also describes well the position of the triple point in the full model with kinetic term \cite{Prekrat:2021uos}.
\section{HCIZ effective action}
\label{section3}

The difference between the curvature term and the potential terms in our action is that after the eigenvalue decomposition, $\Phi = U^\dagger \Lambda U$, the similarity matrices do not cancel under the trace and the integration over $[dU]$ cannot be carried out easily.

In our case, for the expected value of a function $\obs(\Lambda)$ that depends only on the eigenvalues of $\Phi$, we can explicitly write
\begin{equation}\label{avg:F}
    \expval{\obs} 
    = 
    \frac{1}{Z}\int [d\Lambda] \, \obs(\Lambda) \, 
    e^{-N\tr\left(- g_2\Phi^2 + g_4\Phi^4\right)}
\int [dU]\,e^{\, g_r N\tr\lr{URU^\dagger \Lambda^2}} \, .
\end{equation}
As shown in Appendix \ref{appendix:HCIZwrong}, the integration over unitary matrices can be performed similarly to \cite{Kanomata:2022pdo} to obtain
\begin{equation}
\expval{\obs}=
\dfrac{
\phantom{\Bigg|}
\displaystyle 
\int [d\Lambda] \, \frac{\Delta^2(\Lambda)}{\Delta(\Lambda^2)} \, \obs(\Lambda) \,
e^{-N\tr(-g_2\Lambda^2 - g_r R\Lambda^2 + g_4 \Lambda^4)}
\phantom{\Bigg|}
}
{
\phantom{\Bigg|}
\displaystyle
\int [d\Lambda] \, \frac{\Delta^2(\Lambda)}{\Delta(\Lambda^2)} \,
e^{-N\tr(-g_2\Lambda^2 - g_r R\Lambda^2 + g_4 \Lambda^4)}
\phantom{\Bigg|}
} \, .    
\end{equation}
Although quite nice---an eigenvalue problem with different eigenvalue \enquote{masses}---this closed-form result seems incompatible with the standard method for solving the eigenvalue distribution equation since $\Delta(\Lambda^2)$ is not always positive and cannot be absorbed into the exponent.

Instead, we will here show how to circumvent this problem by expanding the curvature contribution in \eqref{avg:F} in powers of $g_r$ with the help of the HCIZ integral
\cite{Zuber2012IntroductionTR,Eynard:2015aea}
\begin{equation}
     I = \int\limits_{\mathclap{U(N)}} [dU]\, e^{\,t\tr AUBU^\dag} 
    = \frac{c_N}{t^{N(N-1)/2}}\frac{\det \mathring{e}^{\,t\ket{a}\bra{\hspace{1pt}b\hspace{1pt}}}}{\Delta(A)\Delta(B)} \, .
    \label{HCIZ}
\end{equation}
In $I$, $A$ and $B$ are Hermitian matrices whose respective eigenvalues $a_i$ and $b_j$ are arranged in vectors
\begin{equation}
    \ket{a} = \col a_i \, ,
    \qquad\quad
    \bra{b} = \row b_j \, ,
\end{equation}
and $\mathring{e}$ is the Hadamard element-wise matrix exponential
\begin{equation}
    \left(\mathring{e}^M\right)_{ij} = e^{M_{ij}},
    \qquad\qquad
    \left(\mathring{e}^{\,t\ket{a}\bra{\hspace{1pt}b\hspace{1pt}}}\right)_{ij} = e^{ta_ib_j}.
\end{equation}
Furthermore, the constant $c_N$ is given by
\begin{equation}
    c_N=\prod_{k=1}^{N-1} k! \, ,
\end{equation}
and $\Delta(A)$ is the Vandermonde determinant
\begin{equation}
    \Delta(A) = \prod_{1 \leq i < j \leq N} (a_j - a_i) \, .
\end{equation}

Due to exponential matrix elements in the numerator of its r.h.s., the HCIZ integral can be written as a power series in $t$
\begin{equation}
     I = \sum_{n=0}^\infty \frac{t^n}{n!}\,I_n \, ,
     \label{t:expansion}
\end{equation}
where the first 4 terms\footnote{We use a simplified notation $\tr^{\mathrlap{n}} \,A := (\tr A)^n$.} are (Appendix \ref{section:proof}):
\begin{subequations}\label{I0123}
\begin{eqnarray}
    I_0 &=& 1 \, ,
    \\ & \notag & 
    \\
    I_1 &=& \frac{\tr A\tr B}{N} \, ,
    \label{trAtrB}
    \\ & \notag & 
    \\
    I_2 &=& \frac{(\tr^{\mathrlap{2}}\, A + \tr A^2)(\tr^{\mathrlap{2}}\, B + \tr B^2)}{2N(N+1)} + \frac{(\tr^{\mathrlap{2}}\, A - \tr A^2)(\tr^{\mathrlap{2}}\, B - \tr B^2)}{2N(N-1)} \, ,\qquad
    \\ & \notag & 
    \\
    I_3 &=& 
    \frac{(\tr^{\mathrlap{3}}\, A + 3\tr A\tr A^2 +2\tr A^3)(\tr^{\mathrlap{3}}\, B + 3\tr B\tr B^2 + 2\tr B^3)}{6N(N+1)(N+2)}
    \notag
    \\
     &&+ \frac{(\tr^{\mathrlap{3}}\, A - 3\tr A\tr A^2 +2\tr A^3)(\tr^{\mathrlap{3}}\, B - 3\tr B\tr B^2 + 2\tr B^3)}{6N(N-1)(N-2)}
    \notag
    \\    
    &&+ \frac{2(\tr^{\mathrlap{3}}\, A - \tr A^3)(\tr^{\mathrlap{3}}\, B - \tr B^3)}{3N(N-1)(N+1)_{\phantom{\big|}}} \, . 
\end{eqnarray}
\end{subequations}
To apply this result to the curvature term \eqref{curvature term}, we set $A = \diag (1,\ldots,N)$ and $B=\Lambda^2$. 

\noindent
On the other hand, $I$ contributes to the effective action as
\begin{align}
    I = \exp\smash{\lr{-\sum_{n=1}^\infty \frac{t^n}{n!}S_n}}
    &= 1 - tS_1 + \frac{t^2}{2!}(S_1^2 - S_2) 
    \nonumber
    \\    
    &-  \frac{t^3}{3!}(S_1^3 - 3S_1S_2 + S_3) 
    \nonumber
    \\
    &+ \frac{t^4}{4!}(S_1^4 - 6S_1^2 S_2 + 3S_2^2 + 4S_1S_3 - S_4)
    + O(t^5) \, ,_{\phantom{\Big|}}
    \label{S:expansion}
\end{align}
so by equating terms in \eqref{t:expansion} and \eqref{S:expansion}, and using (\ref{I0123}a--d), we get  
\begin{equation}\label{S123}
S_1 = -\frac{N+1}{2}\tr\Lambda^2 \, ,
\qquad\quad
S_2 = \frac{1}{12}\tr^{\mathrlap{2}}\,\Lambda^2 - \frac{N}{12}\tr\Lambda^4 \, ,
\qquad\quad
S_3 = 0 \, .
\end{equation}
This means that for \eqref{HCIZ} we can write
\begin{equation}
    I=\exp\lr{-\lr{t S_1+\frac{t^2}{2!} S_2 + \frac{t^3}{3!} S_3}},
\end{equation}
valid up to $O(t^3)$.
Since the higher terms satisfy
\begin{equation}
    I_n = -S_n + \func(S_1,S_2,\ldots,S_{n-1}) \, ,
\end{equation}
all $S_n$'s exist and can be expressed in terms of $I_n$'s, by induction.

An alternative to the above approach is to first expand the integrand in \eqref{HCIZ} in powers of $t$ and then perform term-by-term $[dU]$ integration using group theory techniques. This was first done in the context of fuzzy field theories in \cite{OConnor:2007ibg}. Explicitly, we have 
\begin{align}
    I = 1 
    + 
    t \int\limits_{\mathclap{U(N)}}[dU]\,\tr AUBU^\dag
    +
    \frac{t^2}{2!} \int\limits_{\mathclap{U(N)}}[dU]\,\tr^{\mathrlap{2}} \,AUBU^\dag
    + \ldots \, .
\end{align}
We will thus need to compute integrals of the form
\begin{align}
    \int\limits_{\mathclap{U(N)}} [dU]\, \underbrace{\tr AUBU^\dag\times\ldots\times\tr AUBU^\dag}_{n} 
    \, .
\end{align}
The trick is to rewrite this as a trace in the $n$-fold tensor product
\begin{align}
    I_n=\int\limits_{\mathclap{U(N)}} [dU]\, \tr\left[(A\otimes\ldots\otimes A)(U\otimes\ldots\otimes U)(B\otimes\ldots\otimes B)(U^\dagger\otimes\ldots\otimes U^\dagger)\right]\label{In}
\end{align}
and express the reducible product of $U$'s as a sum of the irreducible representations of $U(N)$. We then have an orthogonality relation
\begin{align}
    \int\limits_{\mathclap{U(N)}} [dU]\,\rho^{\phantom{\dagger}}_{ij}(U)\,\rho^\dagger_{kl}(U)=\frac{1}{\dim\rho}\,\delta_{ij}\delta_{kl}
\end{align}
for a given irreducible representation $\rho$ of $U(N)$. After all the dust settles, we have
\begin{align}
    I_n=\sum_\rho \frac{1}{\dim\rho}\tr_\rho(A\otimes\ldots\otimes A)\tr_\rho(B\otimes\ldots\otimes B)\, ,
\end{align}
where the sum goes over all irreducible representations included in the decomposition of the $n$-fold tensor product of the fundamental representation and $\tr_\rho$ is the trace in the given representation, which in this case simplifies to the character of the respective matrix. Formulas needed for the calculation of the $n=1,2,4$ version of \eqref{In} can be found in \cite{OConnor:2007ibg} and in notation \eqref{S:expansion} confirm the formulas \eqref{S123}. The expression for $S_4$ can be then found to be
\begin{equation}\label{S4}
S_4 = \frac{1}{120}\lr{ N\tr\Lambda^8 + 3\tr^{\mathrlap{2}}\,\Lambda^4 - 4\tr\Lambda^6\tr\Lambda^2 }.
\end{equation}

Finally, by substituting $t \to -16g_r$ and adding $t^nS_n/n!$ terms to the PP action, we obtain the large-$N$ effective action up to $O(g_r^4)$: 
\begin{empheq}[box=\fbox]{multline}
    \label{SRPP}
    S(\Lambda) = N\tr\lr{
    - \lr{g_2 - 8g_r}\Lambda^2 
    + \lr{g_4 - \frac{32}{3}g_r^2}\Lambda^4 
    + \frac{1024}{45}g_r^4\tr\Lambda^8 
    }+
    \\
    + \frac{32}{3}g_r^2\tr^{\mathrlap{2}}\, \Lambda^2
    + \frac{1024}{15}g_r^4\tr^{\mathrlap{2}}\,\Lambda^4
    - \frac{4096}{45}g_r^4\tr\Lambda^6\tr\Lambda^2
    - \log\Delta^2(\Lambda) \, .
\end{empheq}
Note that the curvature term not only shifts the mass and interaction coefficients but also produces new multitrace terms.  

This action will take center stage for the rest of the paper, and we will describe the phase structure of the resulting matrix model both analytically and numerically, comparing the two approaches at the end. 

\section{Analytical phase diagram}\label{section4}

After setting up the matrix model \eqref{SRPP}, which approximates the action \eqref{matrix model} including the effect of the curvature term, let us now proceed to the determination of the transition line of this model. We will first deal with the simple part, essentially not taking into account the $S_4$ term \eqref{S4}, and only use the full action later on.

\subsection{Transition line at third order}

The action \eqref{SRPP} is a multitrace matrix model, which can be solved in the large-$N$ limit by treating it as an effective pure potential model\footnote{For a quick review see \cite{Subjakova:2020prh}.}. Let us first consider only terms up to $O(g_r^3)$, i.e.
\begin{align}
    S(\Lambda) = N\tr\lr{
    -\lr{g_2 - 8g_r}\Lambda^2 + \lr{g_4 - \frac{32}{3}g_r^2}\Lambda^4
    }
    +
    \frac{32}{3}g_r^2\tr^{\mathrlap{2}}\, \Lambda^2
    -\log\Delta^2(\Lambda)
    \, .
\end{align}
The saddle point equation for this model reads\footnote{Using $\partial\lr{ \tr^{\mathrlap{n}}\,\Lambda^m}/\partial \lambda_i= n\tr^{n-1}\Lambda^m \cdot m \lambda_i^{m-1} = mn\,\lambda_i^{m-1}\tr^{n-1}\Lambda^m.$}
\begin{align}
    -2\lr{g_2 - 8g_r
    -\frac{64}{\,3N\,} g_r^2 \tr \Lambda^2}\lambda_i
    + 4 \left(g_4 - \frac{32}{3}g_r^2\right)\lambda_i^3
    =\frac{2}{N}\sum_{i\neq j}\frac{1}{\lambda_i-\lambda_j} \, .
\end{align}
If we, in the large-$N$ limit, replace $\tr\Lambda^2$ by its expectation value $\expval{\tr \Lambda^2}$, the above is a saddle point equation for an effective pure potential model with shifted parameters
\begin{subequations}\label{effcs}
\begin{align}
g_{2,\text{eff}} &= g_2 - 8g_r-\frac{64}{\,3N\,} g_r^2 \avg{ \tr \Lambda^2},
\\
g_{4,\text{eff}} &= g_4 - \frac{32}{3}g_r^2 \, .
\end{align}
\end{subequations}
We could, in principle, solve this effective model, but with the expression for the second moment of the distribution
\begin{equation}
    m_2=\frac{\avg{\tr \Lambda^2}}{N} = \int d\lambda \; \rho(\lambda) \, \lambda^2
    \label{self-consistency}
\end{equation}
viewed as a new self-consistency condition\footnote{Note that the distribution $\rho$ depends on $g_{\text{eff}}$'s and thus on $\avg{\tr \Lambda^2}$.}. Here we, however, need even less since we are interested only in the phase transition line
\begin{align}
    g_{2,\textrm{eff}} = 2\sqrt{g_{4,\textrm{eff}}} \, , 
\end{align}
along which it holds
\begin{align}
    \frac{\avg{\tr \Lambda^2}}{N} = \frac{1}{\sqrt{g_{4,\textrm{eff}}}} \, .
\end{align}
Substituting the effective parameters from (\ref{effcs}a--b) leads to
\begin{align}\label{full 3rd order transition line}
    g_2 = 
    2\sqrt{g_4 - \frac{32}{3}g_r^2} + 8g_r + \frac{\displaystyle \frac{64}{3}g_r^2}{\displaystyle \sqrt{g_4  - \frac{32}{3}g_r^2}} 
    \, .
\end{align}

\begin{figure}[t]
\centering  
\includegraphics[width=0.80\textwidth]{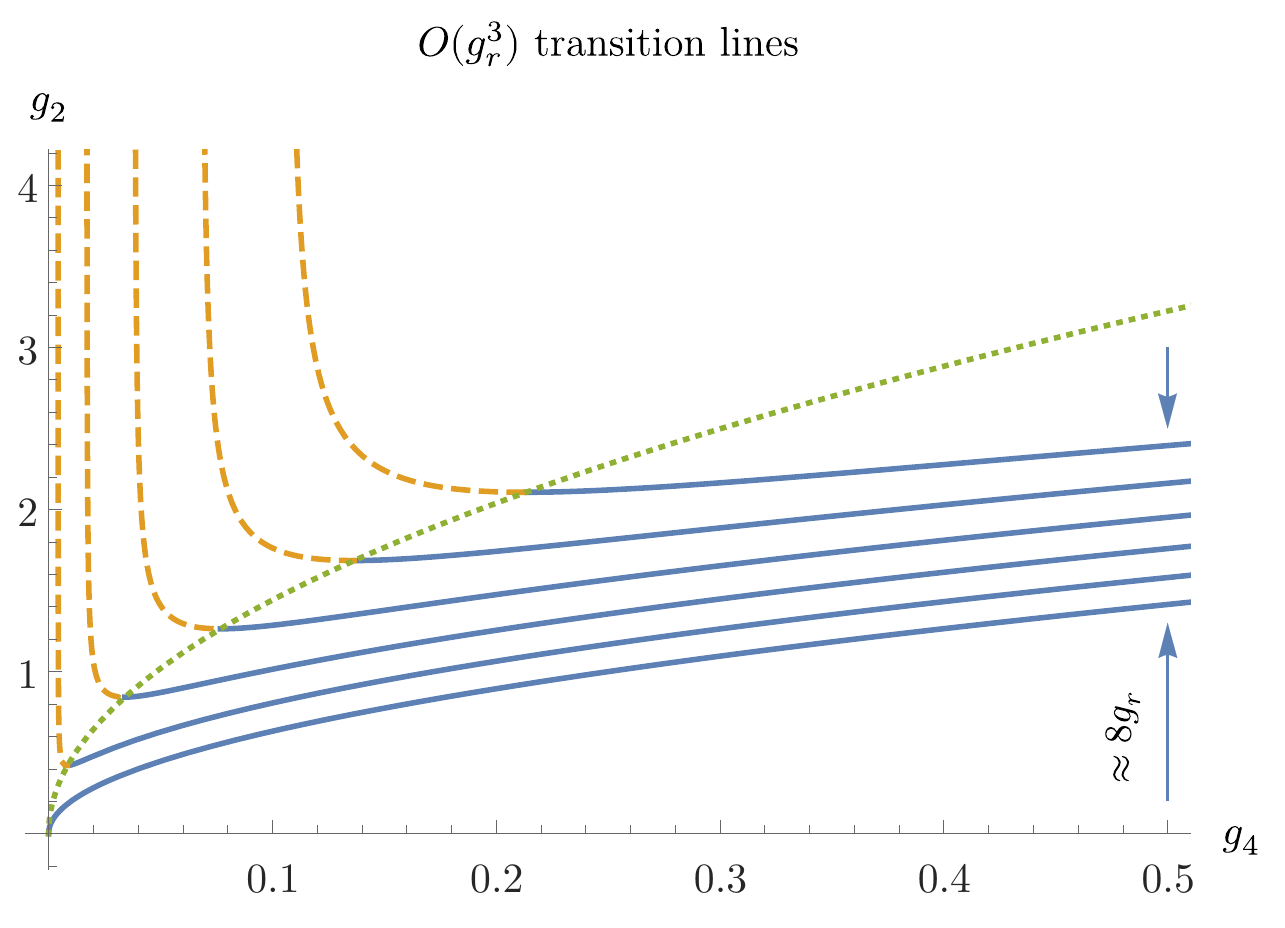}
\caption{
Plots of the $O(g_r^3)$ transition line \eqref{3rd order transition line} for values $g_r\in\{0.00,\,0.02,\,0.04,\,0.06,\,0.08,\,0.10\}$ (bottom to top), where the dashed, orange part is not to be trusted. The dotted, green line is the locus of transition lines' minima $(g_4,\,g_2) = (64 g_r^2/3,\,8(1+\sqrt{3})g_r)$.
}
\label{figure:O3}
\end{figure}

There are several good reasons not to trust this line all the way to ${g_4=0}$. First of all, we only took the action up to $O(g_r^3)$. Expanding the transition line equation to the same precision gives
\begin{align}\label{3rd order transition line}
    g_2 = 
    2\sqrt{g_4 }+8g_r+\frac{32}{3}\frac{g_r^2}{\sqrt{g_4 }}
    + O(g_r^4)
    \, .
\end{align}
Plots of this transition line for various values of $g_r$ are shown in Figure \ref{figure:O3}. Also, both analytical arguments behind \eqref{16gr} and numerical results of Section \ref{section:numerical} imply that the transition line starts near $(g_4,\,g_2) = (0,\,16 g_r)$.

But we also need to keep in mind that the approximation breaks also for large moments of the distribution $m_2$, which means that it is valid only for reasonably large values of $g_4$. The minimum of the curve \eqref{full 3rd order transition line} lies at 
\begin{equation}\label{O3min}
    g_4 = \frac{64g_r^2}{3} \approx 21g_r^2 \, ,
    \qquad\quad
    g_2 = 8\lr{1 + 2\sqrt{\frac{2}{3}}}g_r \approx 21g_r \, ,
\end{equation}
and we expect the expression not to be trusted for smaller values of $g_4$.

\subsection{Transition line at fourth order}

\begin{figure}[t]
\centering  
\includegraphics[width=0.80\textwidth]{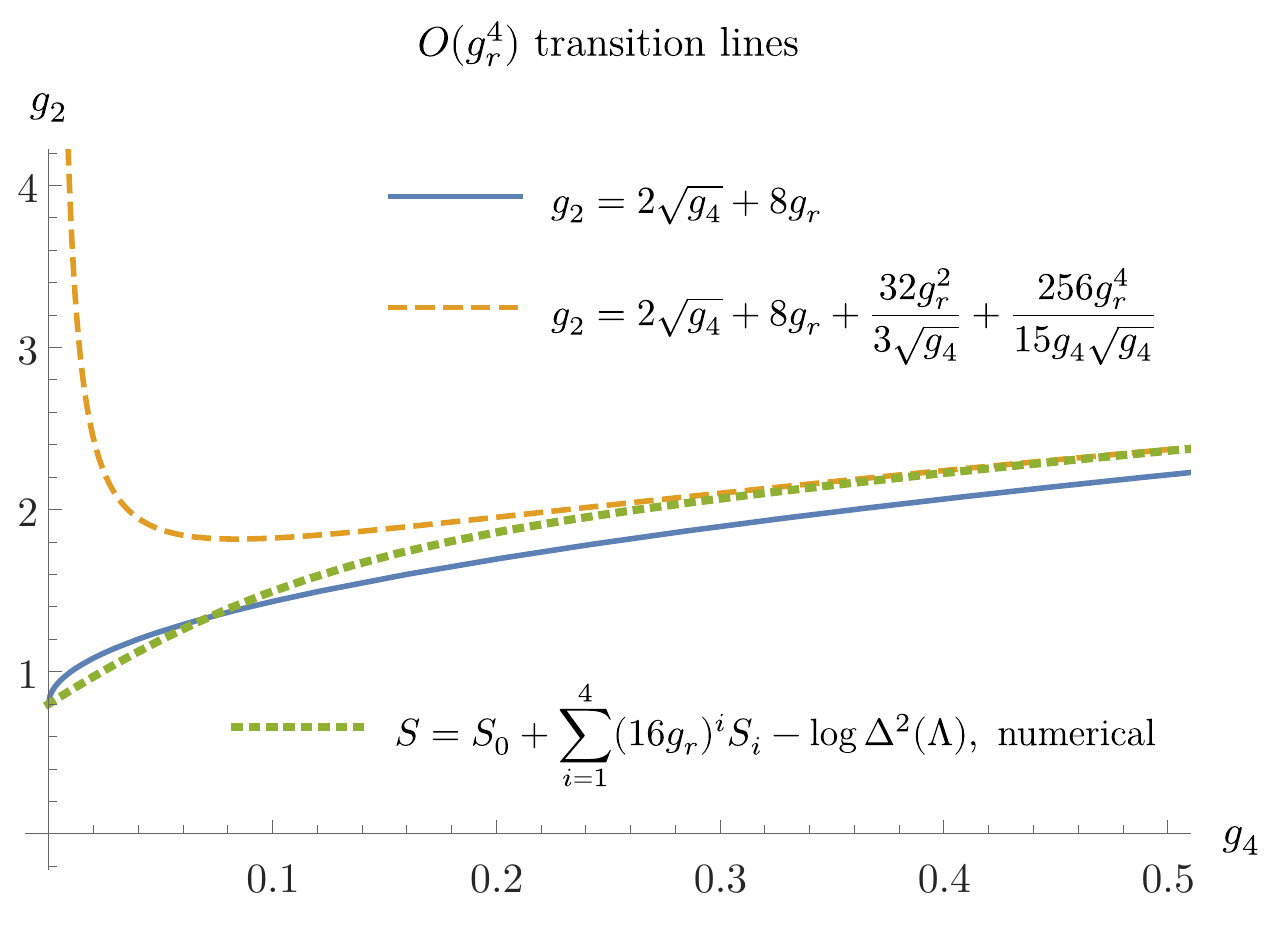}
\caption{Blue solid line represents $O(g_r)$ correction from \eqref{3rd order transition line}. Orange dashed line represents $O(g_r^4)$ transition line \eqref{4th order transition line}. Green dotted line represents numerically calculated transition line for the multitrace model \eqref{SRPP} (see appendix \ref{appendix:4th order distribution}). All three plots are for $g_r = 0.1$.}
\label{figure:O4}
\end{figure}

Contrary to the third order action, if we include the $O(g_r^4)$ term \eqref{S4}, the effective single trace model is not a simple quartic model and becomes a polynomial model of the sixth order. The derivation of the eigenvalue distribution equation for \eqref{SRPP} and the transition line between the 1-cut-sym and the 2-cut-sym solution is left for Appendix \ref{appendix:4th order distribution}. It yields the following transition line equation
\begin{equation}
    \label{4th order transition line}
    \boxed{
    g_2 = 2\sqrt{g_4} + 8g_r + \frac{32}{3}\frac{g_r^2}{\sqrt{g_4}} + \frac{256}{15}\frac{g_r^4}{g_4\sqrt{g_4}} \, ,
    }
\end{equation}
which agrees with the $O(g_r^3)$ results in \eqref{3rd order transition line}. This function has a minimum at
\begin{equation}
    g_4 = \frac{8}{15}\lr{\sqrt{115}+5}g_r^2 \approx 8.4 g_r^2 \, ,
    \qquad\quad
    g_2 \approx 18g_r \, ,
\end{equation}
and for smaller values of $g_4$ quickly blows up to infinity, as shown in Figure \ref{figure:O4}. 
This is again at odds with the expectation from Section \ref{section:GW} that the transition line starts at the $(g_4,\,g_2)=(0,\,16g_r)$. However, the minimum of the $O(g_r^4)$ line is closer to this point than the minimum of the $O(g_r^3)$ line \eqref{O3min} and this trend would presumably persist for higher-order terms.

We will, therefore, try to guess the non-perturbative expression which expands into \eqref{4th order transition line} for large $g_4$. As it is demonstrated in \cite{Polychronakos:2013nca,Tekel:2017nzf}, the kinetic term on the fuzzy sphere deforms the square-root behaviour of the PP model's transition line into 
\begin{equation}
    2g_2 = 5\sqrt{g_4} + \frac{1}{1-\exp(1/\sqrt{g_4})} \, .
\end{equation}
The additional term on the r.h.s. is clearly non-analytical in $g_4$. If we think of the curvature as an effective kinetic operator, we can expect a similar deformation in our model. The ansatz
\begin{equation}
    g_2 = \alpha_1\sqrt{g_4} + \frac{\alpha_2\,g_r}{1 - \exp(\alpha_3\,g_r/\sqrt{g_4})} 
\end{equation}
would correct the transition line's erratic behaviour for $g_4 \to 0$, but its expansion is unfortunately incompatible with \eqref{4th order transition line}. A simple expression
\begin{equation}
    g_2 = \frac{16g_r}{1-\exp(-8g_r/\sqrt{g_4})} \, ,
\end{equation}
however, fits well with the simulation data in the Figure \ref{figure:N24} of Section \ref{section:numerical}, and reproduces theoretical prediction up to $O(g_r^3)$, but fails at $O(g_r^4)$:
\begin{equation}
    2\sqrt{g_4} + 8g_r + \frac{32}{3}\frac{g_r^2}{\sqrt{g_4}} - \frac{512}{45}\frac{g_r^4}{g_4\sqrt{g_4}} + O(g_r^6) \, .
\end{equation}
This could be corrected, for example, by an additional term\footnote{The term is introduced by hand after a series of guesses.}
\begin{equation}
    \alpha\sqrt{g_4}\cdot\lr{1-\exp(-(\beta g_r/\sqrt{g_4})^n)}^{4/n} = \frac{\alpha\beta^4\, g_r^4}{g_4\sqrt{g_4}} + \ldots \, , 
    \qquad
    \alpha\beta^4 = \frac{256}{9} \, ,
\end{equation}
which has a nice property that does not move the starting point $(g_4,\,g_2)=(0,\,16g_r)$ of the transition line.

\section{Numerical phase diagram}
\label{section:numerical}

Let us now compare our analytical predictions against the outcomes of numerical simulations using the Hamiltonian Monte Carlo method. They are especially important, since they provide insight into complete, nonperturbative results.

\begin{figure}[b]
\centering 
\includegraphics[width=1.00\textwidth]{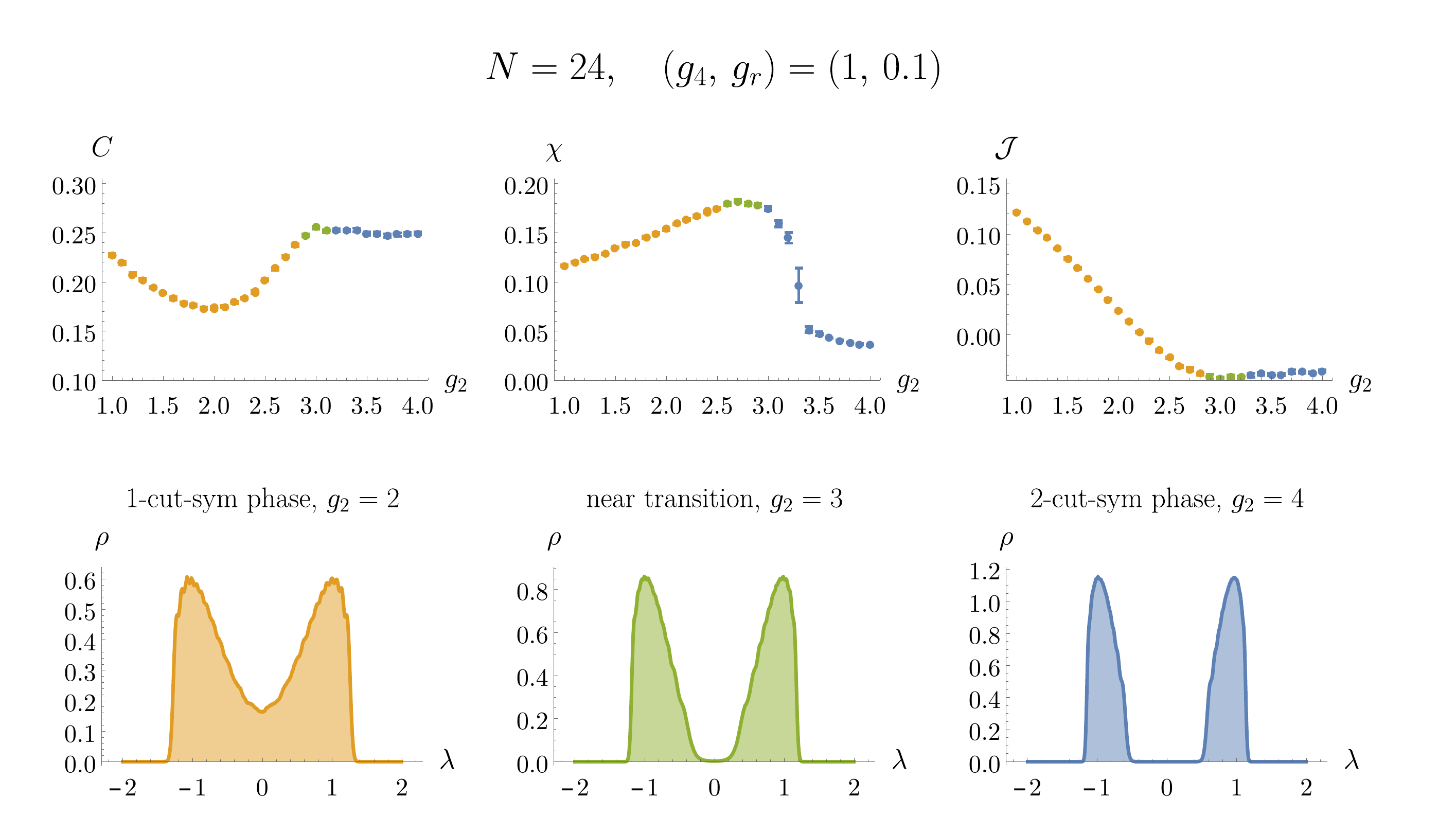}
\caption{
Transition point detection from peaks and edges of $C$, $\chi$, and $\mathcal{J}$ (top), and from topology change of the eigenvalue-distribution support (bottom).}
\label{figure:peaks}
\end{figure}

To study the system numerically, we utilized the ordinary Hamiltonian Monte Carlo method \cite{handbookMCMC}, occasionally verifying the results by the eigenvalue-flipping procedure \cite{Kovacik:2022kfh} enhanced by a parallel computing tool \cite{tange_ole_2018_1146014}. The basic idea is that the system is initialized in a random configuration and then undergoes a series of changes that resemble random external impulses. The system moves across the configurational space in a manner with the same statistical properties as those given by the partition sum and we measure it periodically to obtain the mean values of observables of interest. The calibration of the methods was done on the PP model, for which both reproduced the well-known results. For more details about the algorithms, we refer the interested reader to \cite{Prekrat:2020ptq,Kovacik:2022kfh}.

We are interested in the phase transition from the 1-cut-sym to the 2-cut-sym phase. There are many observables that are suitable for identifying this transition, and while they might not completely agree for finite values of $N$, they converge in the large-$N$ limit, as demonstrated in \cite{Prekrat:2020ptq,Prekrat:2021uos}.

Two popular choices in the literature are the susceptibility, 
\begin{equation}
    \chi  = \frac{\Var \abs{\tr \Phi}}{N} \, ,
\end{equation}
and the specific heat,
\begin{equation}
    C = \frac{\Var S}{N^2} \, .
\end{equation}
Another option is to use the following combination
\begin{equation}\label{order:J}
    {\cal{J}} = - \frac{\tr \Phi^4}{N} + \frac{\tr^{\mathrlap{2}}\; \Phi^2}{N^2}+ \frac{1}{4} \, .
\end{equation}
In the PP case, this quantity vanishes in the 2-cut regime, while behaving like $(g_2-g_2^*)^2/32$ close to the phase transition point $g_2^*$ in the 1-cut regime.

\begin{figure}[t]
\centering  
\includegraphics[width=0.80\textwidth]{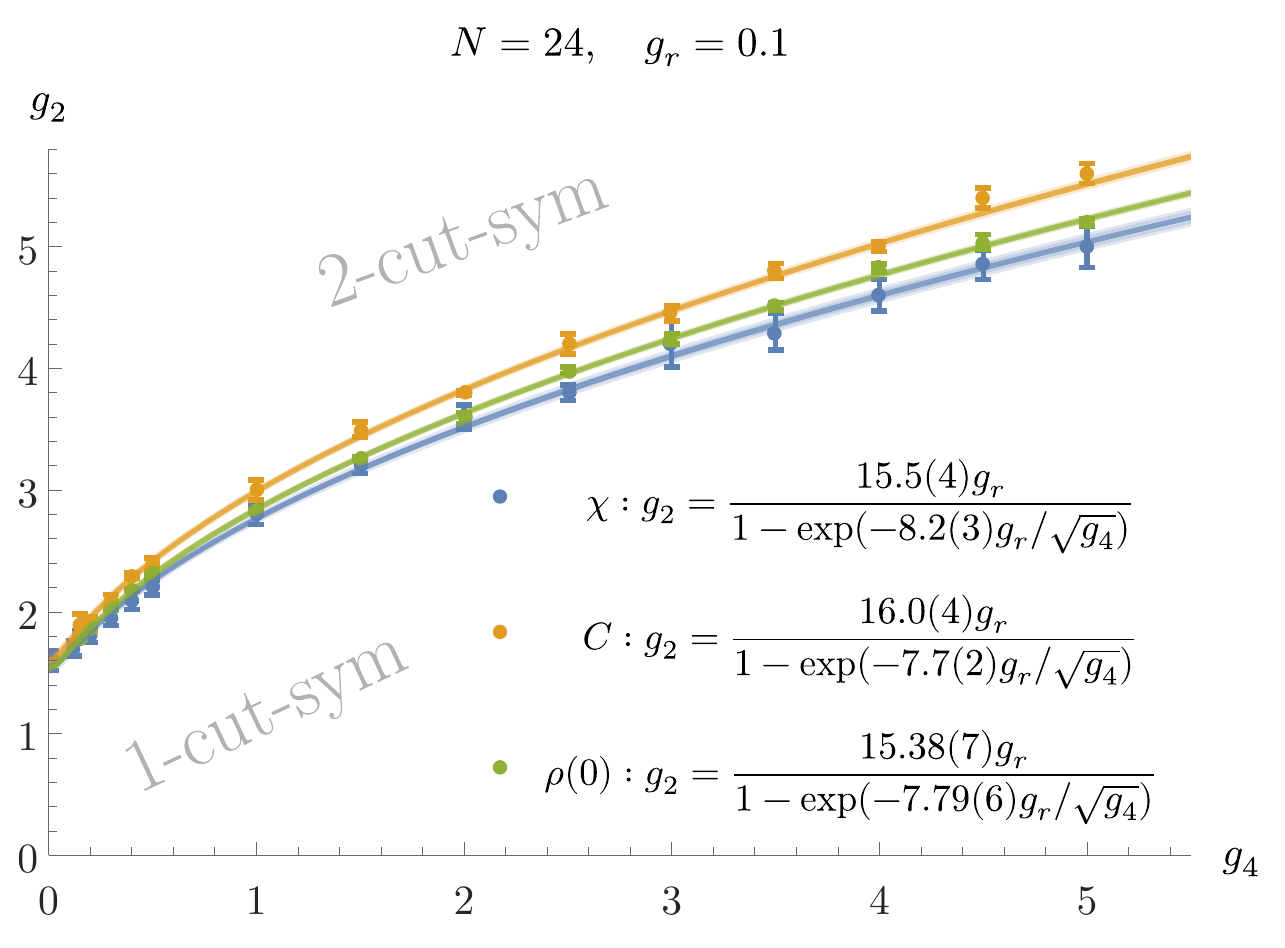}
\caption{
Hybrid Monte Carlo phase diagram constructed from peaks in $C$ and $\chi$, and from splitting points of eigenvalue distribution $\rho(0)=0$. A simple function $g_2 = 16g_r\big/\lr{1-\exp\lr{-8g_r/\sqrt{g_4}}}$ is a good approximation for the numerical transition lines and correctly reproduces theoretical expression up to $O(g_r^3)$.  
}
\label{figure:N24}
\end{figure}

\begin{figure}[t]
\centering  
\includegraphics[width=0.8\textwidth]{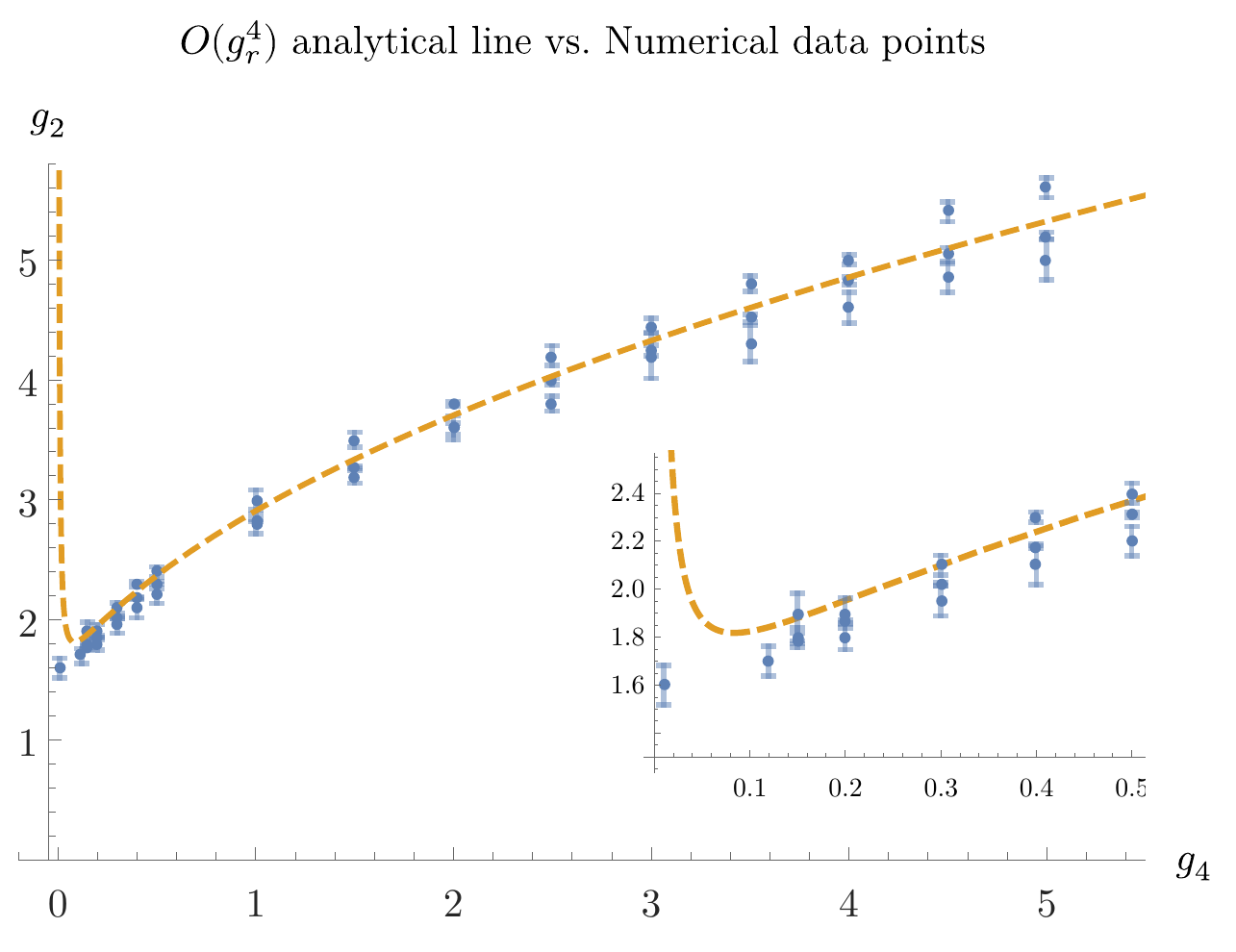}
\caption{
Analytical $O(g_r^4)$ transition line \eqref{4th order transition line} vs. $N=24$ numerical data points. All the data points from $C$, $\chi$ and $\rho(0)$ are now the same color.  
}
\label{figure:THvsNUM}
\end{figure}

Figure \ref{figure:peaks} shows examples of these three observables. Peaks and edges in their profiles indicate the points of phase transition. The figure also confirms that the phase transition is connected to the change in topology of eigenvalue-distribution support. Using any of these observables, one can construct the phase diagram by varying the values of $g_4$. In Figure \ref{figure:N24}, we see the one obtained for $N=24$ and $g_r=0.1$. The numerical phase diagram shows good agreement with the analytical approximations obtained in previous sections, as demonstrated in Figure \ref{figure:THvsNUM}.

We can additionally define an observable that directly relates to the eigenvalue distribution. The observable 
\begin{equation}\label{Pep}
P_\varepsilon = \int \limits_{\mathclap{-\varepsilon/2}}^{\mathclap{+\varepsilon/2}} \rho(\lambda) \, d\lambda 
\end{equation}
is easily obtained from the eigenvalue histograms collected during the simulation, and it 
measures the gap in the eigenvalue distribution. Its limit $P_{\varepsilon \to 0}$ is zero in the case of the gaped 2-cut distribution and non-zero for 1-cut distributions. Since 
\begin{equation}
    P_\varepsilon \approx \varepsilon\rho(0) + O(\varepsilon^2) \, ,
\end{equation}
$P_\varepsilon/\varepsilon$ is our best estimate for the value of $\rho(0)$ that captures the transition in a transparent way.

Let us look at the center of the PP eigenvalue distribution \cite{Filev:2014jxa}
\begin{equation}
    \rho_\text{\smaller PP}(0) = \frac{\sqrt{g_2^2+12 g_4}-2 g_2}{3 \pi }\sqrt{\frac{\sqrt{g_2^2+12 g_4}+g_2}{3 g_4}} \, .
\end{equation}
Near the transition point $g_2^* = 2\sqrt{g_4}$, it can be expanded into
\begin{equation}
    \label{rho0}
    \rho_\text{\smaller PP}(0) = -\frac{\sqrt[4]{4g_4}}{\pi}
    \cdot
    \lr{
    \delta
    +
    \frac{\delta^2}{8}
    -
    \frac{\delta^4}{512}
    }
    +
    O(\delta^5) \, ,
    \qquad\quad
    \delta = \frac{g_2 - g_2^*}{g_2^*} \, .
\end{equation}

\begin{figure}[t]
\centering 
\includegraphics[width=1.00\textwidth]{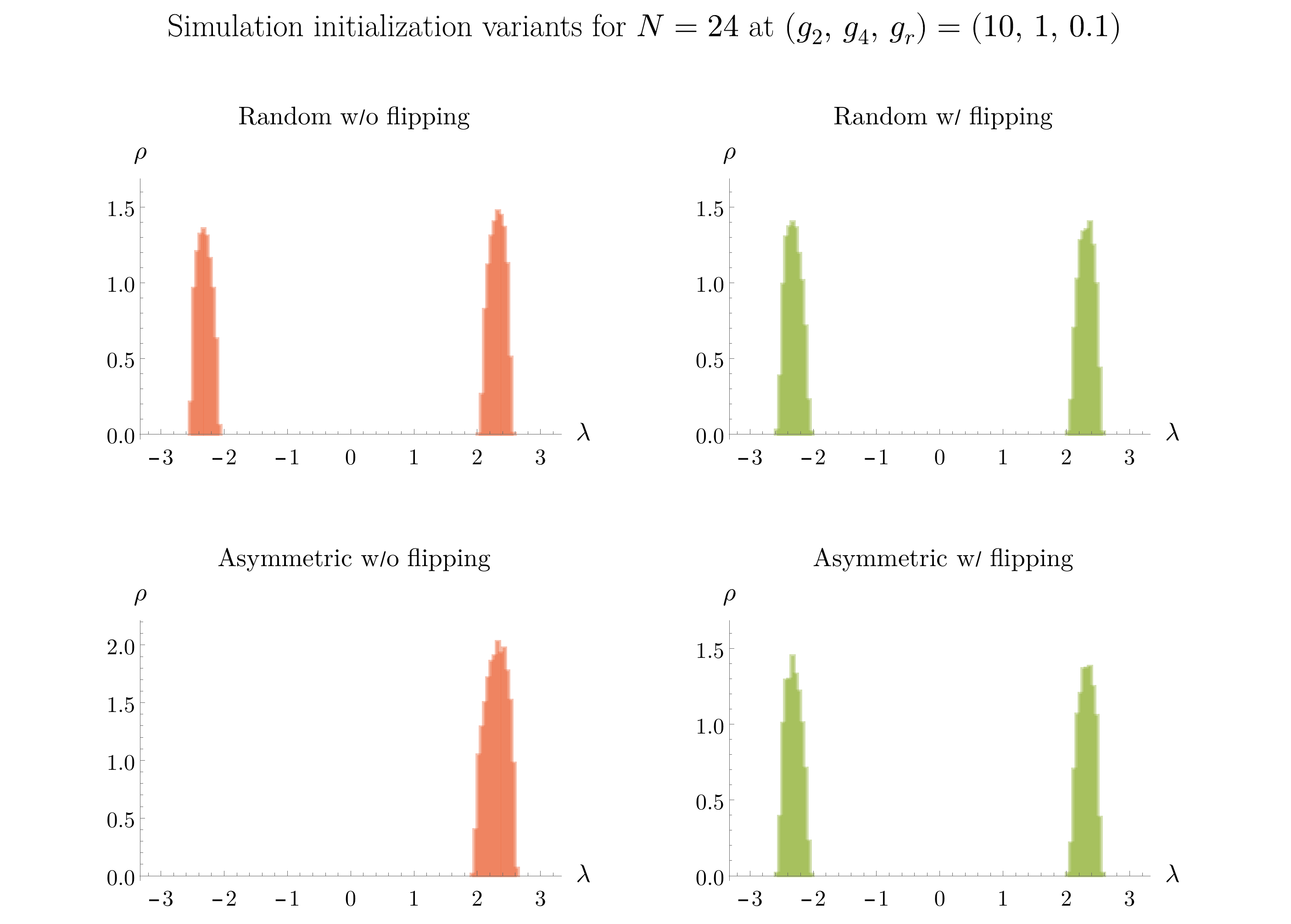}
\caption{
Eigenvalue-distribution histograms obtained from simulations that were either initialized from random configuration or with all eigenvalues in the same potential well (1-cut-asym). The eigenvalue-flipping procedure \cite{Kovacik:2022kfh} was either turned on or off. It can be observed that the 1-cut-asym phase is only apparent and vanishes if the system is allowed to tunnel efficiently to the energetically preferred state.}
\label{figure:ordered}
\end{figure}

We see that a linear behaviour with a slight quadratic correction is an excellent approximation for $\rho(0)$ even farther away from the transition point, making it easy to locate. Repeating the analysis for the $O(g_r^3)$ action with the help of Appendix \ref{appendix:4th order distribution}, the previous formula is deformed into
\begin{equation}
    \label{rho0order2}
    \rho(0) = -\frac{\sqrt[4]{4g_4}}{\pi}
    \lr{1 + \frac{4g_r}{\sqrt{g_4}} - \frac{8g_r^2}{3g_4}}
    \cdot
    \lr{
    \delta
    +
    \frac{1}{8}\lr{1 + \frac{4g_r}{\sqrt{g_4}} - \frac{128g_r^2}{3g_4}}\delta^2
    }
    +
    O(\delta^3) \, .
\end{equation}
As shown for $(g_r,\,g_4) = (0.1, \,1)$ in Appendix \ref{appendix:extrapolation}, these coefficients agree very nicely with the large-$N$ extrapolated numerical results:
\begin{subequations}
\begin{align}
\text{theoretical:} \qquad \rho(0) &\approx -0.618\cdot\lr{\delta + 0.122\cdot\delta^2},
\\
\text{numerical:} \qquad \rho(0) &\approx -0.611(5)\cdot\lr{\delta + 0.134(5)\cdot\delta^2}.
\end{align}
\end{subequations}
Better yet, the extrapolated triple-point position matches the theoretical prediction from \eqref{3rd order transition line} to less than a half of a percent:
\begin{subequations}
\begin{align}
\text{theoretical:} \qquad  g_2^* &\approx 2.908 \, ,
\\
\text{numerical:} \qquad  g_2^* &\approx 2.912(9) \, .
\end{align}
\end{subequations}

These high-precision results, however, come with a warning. The issue with numerical methods is that the system can be stuck in a false vacuum. For example, we can choose the values of $g_2$ and $g_4$ so the potential has two deep potential wells and that the system prefers to be in the 2-cut-sym phase. If we, instead, initialize it with the 1-cut-asym solution, the tunneling process to the correct vacuum can take a staggeringly long time. In these cases, the eigenvalue-flipping procedure can significantly shorten the thermalization process. We have used this procedure to confirm that our model has no 1-cut-asym phase in its phase diagram (Figure \ref{figure:ordered}), unlike the full model with kinetic term \cite{Prekrat:2021uos}.

We are now ready to see how well our analytical predictions model the full action with the kinetic term \eqref{GW matrix model}. The results of simulations for $N=24$ and $g_r = 0.1$ in the strong-interaction regime
\begin{eqnarray}
    g_4 \geq 1 \gg 32g_r^2/3 \approx 0.11 \, ,
\end{eqnarray}
are collected in Table \ref{table:N24}. The $g_r = 0.1$ lines are fitted through data points used for phase diagrams in \cite{Prekrat:2021uos}; this is also where the $g_r = 0$ expressions come from. As a very important and somewhat unexpected observation note, even with the kinetic term, the shift between simulated transition lines for the model with and without the curvature term agrees nicely with the value $8g_r$ expected from the $O(g_r)$ contribution in \eqref{3rd order transition line}.

\renewcommand{\arraystretch}{1.55}
\begin{table}[t]
\caption{The full model \eqref{GW matrix model} transition line fits for $N=24$ and $g_4 \geq 1$.}
\centering
\setlength{\tabcolsep}{1em}
\begin{tabular}{ccccc}
    \hline
    transition & \quad 1-cut-sym/2-cut-sym &  \quad 2-cut-sym/1-cut-asym
    \\
    \hline
    $g_r=0.0$  & \quad $g_2=2.67(5)\sqrt{g_4}-0.55(7)$  & \quad $g_2=3.99(4)\sqrt{g_4}-0.90(5)$ \\
    $g_r=0.1$ & \quad $g_2=2.60(4)\sqrt{g_4}+0.32(6)$ & \quad $g_2=3.95(6)\sqrt{g_4}-0.3(1)\phantom{0}$ 
    \\ 
    \hline
    shift & \quad $0.87(9) = 8.7(9)g_r$ & \quad $0.6(1) = 6(1)g_r$
    \\
    expected & \quad $8g_r + O(g_r^2)$ & \quad $8g_r  + O(g_r^2)$
    \\
    \hline
\end{tabular}
\label{table:N24}
\vspace{10pt}
\end{table}

This can be partially explained by looking at the classical action. Namely, for $g_2 > 16 g_r$, the 1-cut-asym EOM solution can be expanded as 
\begin{equation}
    \Phi = \sqrt{\frac{g_2\1 + g_r R}{2g_4}} = \sqrt{\frac{g_2}{2g_4}}\lr{\1 + \frac{g_r R}{2g_2}} + O(g_r^2) \, .
\end{equation}
Since $\K\1=0$, the kinetic term reads
\begin{equation}
    \tr\Phi\K\Phi = \frac{g_r}{2g_2}\sqrt{\frac{g_2}{2g_4}}\tr\K R + O(g_r^2) \, . 
\end{equation}
But $\K$ is just a commutator, so its trace vanishes giving
\begin{equation}
    \tr\Phi\K\Phi = 0 + O(g_r^2) \, , 
\end{equation}
causing the $O(g_r)$ effects from the curvature term to dominate and produce the $8g_r$ shift also in the 1-cut-asym regime. 

This also explains the numerical $\sqrt{g_4}$-coefficients for the 2-cut-sym/1-cut-asym transition lines in Table \ref{table:N24}. With the kinetic term gone, they are near the PP model value $\sqrt{15} \approx 3.87$ \cite{Tekel:2017nzf,Subjakova:2020shi}. For the same reason, 1-cut-sym/2-cut-sym transition line \eqref{4th order transition line} starts with $2\sqrt{g_4}$.

The above analysis of the kinetic term fails in the striped phase, where eigenvalues can take alternating signs. However, since simulations imply that kinetic term $\expval{\tr \Phi\K\Phi}$ stays fairly constant throughout the striped phase (Figure 1 in \cite{Prekrat:2020ptq}), we can simply disregard its contribution---a constant does not affect the action derivative which gives the eigenvalue distribution equation and therefore the phase transition position. This constant, however, lifts the energy of the striped phase compared to the ordered phase, allowing the 1-cut-asym solution to manifest itself.

\section{Conclusions and discussion}

In the presented work, we have studied the consequences of the curvature term in the matrix formulation of the Grosse-Wulkenhaar model, both analytically and numerically. We have approximated the effect of the curvature term by a multitrace matrix model derived by perturbative integration of the angular degrees of freedom. The numerical part consisted of the Hamiltonian Monte Carlo calculation of the full matrix integral, incorporating the eigenvalue-flipping algorithm \cite{Kovacik:2022kfh} and using two new order parameters (\ref{order:J},\,\ref{Pep}). As a result, we have obtained the phase diagram of the model, which includes the disordered and non-uniformly ordered phases---both preserving the $\Phi\to-\Phi$ symmetry of the matrix action---and a phase transition line between the two. This has been done in various ways, which all agree.

We confirmed the curvature-induced transition line shift which was a key ingredient for the removal of the striped phase in \cite{Prekrat:2021uos}, at least in the strong-interaction regime. However, the desired analytical result in the weak-interaction regime is still out of reach. A similar result should hold in the 4-dim version of the GW model \cite{Grosse:2004yu} where the field is coupled to a direct sum of two 2-dim curvatures \cite{Buric:2009ss}.

To start the discussion, let us address the elephant in the room---the asymmetric solutions that have been observed in the previous numerical studies \cite{Prekrat:2020ptq,Prekrat:2021uos} of the full GW model \eqref{GW matrix model}. Clearly, the multitrace approximation \eqref{SRPP} has no chance of supporting a stable asymmetric solution, as it includes only the even moments of the matrix. Moreover, neither does the curvature model $S_{\text{\smaller K}+\text{\smaller PP}}$ \eqref{matrix model}, since the probability distribution is a function of even powers of the eigenvalues, as can be seen from the HCIZ expression \eqref{HCIZ}. This means that the existence of the asymmetric solution must be solely due to the kinetic term. If we rewrite the double commutators in \eqref{kinetic} under the trace as
\begin{align}
    \tr \Phi\comm{X}{\comm{X}{\Phi}} = 2\tr X^2 \Phi^2 - 2\tr \Phi X \Phi X
\end{align}
we can see that the first term can be straightforwardly treated using the same techniques as in this work; however, the second term is more complicated, and the $[dU]$ integral includes four $U$ matrices. It would be interesting to see if this part can be reasonably treated by the character techniques mentioned in Section \ref{section:GW}, which lead to \eqref{S4}.

The techniques of this paper should also be relevant in the gauge model on the truncated Heisenberg algebra \cite{Buric:2016lly}, where one of the gauge field components is coupled to the curvature.

As a possible different conceptual direction for future research, it would be interesting to examine a connection between the GW model and a two-dimensional superfluid vortex crystal in the lowest Landau level studied in \cite{Jeevanesan:2019jty}, due to the similar action and the specific heat profile. This would fit into the tradition of finding effective noncommutativity in condensed matter systems. In this model, our curvature term is mapped to the coupling between the field and the magnetic vector potential.

This connection between models might also have astrophysical consequences. Namely, as shown in \cite{Prekrat:2021uos}, the transition-line shift in our model allows the model's bare parameters to start in different phases depending on the strength of the curvature and its coupling to the matter field. More precisely, in the effectively flat space ($g_r=0$), weakly interacting particles could form the stripped phase, while the presence of curvature ($g_r\neq0$) would confine them to the disordered phase. Also, \eqref{trAtrB} tells us that coupling of any external field to $\Phi^2$ would, in principle, shift the mass parameter and, consequently, the transition lines. In a more realistic NC model, a coupling with a gravitational field of the galaxy ($\Leftrightarrow R\Phi^2$) could thus cause the dark matter ($\Leftrightarrow \Phi$) to behave differently in regions with strong and weak gravity (or on larger scales where the space is effectively flat), which is, as we understand, the idea behind the superfluid dark matter hypothesis \cite{Berezhiani:2015bqa,Hossenfelder:2020yko,Mistele:2022vhh}. If the curvature term is modified to flatten out at large distances, the change in the dark matter's behavior with scale could perhaps be realized as a decoupling of the matrix field into a direct sum of two components of different masses and in different phases.

 There is also room for a more technical expansion of the present work. The derivation of the $S_n$ terms in Appendix \ref{section:proof} systematically increases matrix-element powers in the (modified) Vandermonde determinant. This means that each higher-order term is determined by the terms of the lower order, as in \eqref{recursion}. Perhaps there is a way to extend and solve this recursive equation to all orders, which could help us obtain the non-perturbative result for the transition line. 
 
 As a final remark (and perhaps related to the previous point), let us mention that there has been a very successful program treating correlation functions in a model similar to \eqref{matrix model} resulting in a deep connection to topological recursion theory, see e.g. \cite{Branahl:2020yru,Branahl:2021slr} and references therein. There are a few differences from our work, most notably, the quadratic part had to have only positive eigenvalues, but it definitely would be interesting to see if any of these results are applicable in our case.

\acknowledgments

This research was supported by:
\hspace{20pt}
\begin{itemize}[leftmargin=20pt]
    \item the Ministry of Education, Science and Technological Development, Republic of Serbia:
    \begin{itemize}[leftmargin=12pt]
        \item Grant No. 451-03-68/2022-14/200161, University of Belgrade -- Faculty of Pharmacy,
        \item Grant No. 451-03-68/2022-14/200162, University of Belgrade -- Faculty of Physics,\phantom{....}
    \end{itemize}
    \item VEGA 1/0703/20 grant \emph{Quantum structure of spacetime}. 
\end{itemize}

\appendix
\section{HCIZ formula example}

As an example of the HCIZ formula \eqref{HCIZ}, let us consider the $N=2$ case with $t=1$ and a simple choice of matrices $A=B=\sigma_3$, where $\sigma_3$ is the diagonal Pauli matrix with eigenvalues $\pm 1$.

To evaluate the l.h.s. of \eqref{HCIZ} we need to choose a parametrization for $U(2)$
\begin{equation}
U = \left( \begin{array}{cc}
\;e^{i \left(\varphi_0 + \varphi_1\right)} \cos \theta\;\,
& 
\;\,e^{i\varphi_1} \sin \theta\;
\\ 
\;\mathllap{-}e^{i \left(\varphi_0 + \varphi_2\right)} \sin \theta\;\,
& 
\;\,e^{i \varphi_2} \cos \theta\;
\end{array} \right),
\end{equation}
where $0 \le \varphi \le 2\pi$ and $0 \le \theta \le \pi/2$.
With this, the function in the exponent under integration is

\begin{equation}
\tr AUBU^\dagger = 2 \cos  \varphi_0  \sin 2 \theta 
\end{equation}
and the integration gives

\begin{equation}
\int d \theta \ d \varphi_0 \ d \varphi_1 \ d \varphi_2 \ \frac{\cos \theta \sin \theta}{4 \pi^3} \ e^{2 \cos  \varphi_0  \sin 2 \theta } = 1.81343...\, .
\end{equation}

The $c_N$ factor in \eqref{HCIZ} is in this case simply equal to 1 and both Vandermonde determinants are $\Delta(\sigma_3) = 2$. The Hadamard element-wise matrix exponential gives

\begin{equation}
\left| \begin{array}{cc}
   e  & 1/e \\
  1/e  &  e
\end{array} \right| = 2 \sinh 2 \, .
\end{equation}
After plugging everything together, most of the factors cancel, and the r.h.s. becomes  
\begin{equation}
    \frac{\sinh 2}{2} = 1.81343... \, .
\end{equation}
\section{HCIZ without expansion}
\label{appendix:HCIZwrong}

Let us transform the partition function of the model \eqref{matrix model} by using the HCIZ formula:

\begin{align}
     Z 
     &= \int [d\Phi] \, e^{-N\tr(-g_2\Phi^2 - g_r R\Phi^2 + g_4 \Phi^4)}
     \\
     &= \int [dU] [d\Lambda] \, \Delta^2(\Lambda) \, e^{-N\tr(-g_2\Lambda^2 - g_r RU\Lambda^2U^\dag + g_4 \Lambda^4)}
     \\
     &= \int [d\Lambda] \, \Delta^2(\Lambda) \, e^{-N\tr(-g_2\Lambda^2 + g_4 \Lambda^4)} \int [dU] \, e^{\,Ng_r\tr RU\Lambda^2U^\dag}
     \\
     & \stackrel{\mathclap{\text{\tiny{HCIZ}}}}{=} \int [d\Lambda] \, \Delta^2(\Lambda) \, e^{-N\tr(-g_2\Lambda^2 + g_4 \Lambda^4)} 
     \cdot
     \frac{c_N}{(Ng_r)^{N(N-1)/2}}\frac{\det \mathring{e}^{\,Ng_r\ket{R}\bra{\hspace{1pt}\Lambda^{\smash{2}}\hspace{1pt}}}}{\Delta(R)\Delta(\Lambda^2)}
     \\
     &= \frac{1}{(16g_r)^{N(N-1)/2}} \int [d\Lambda] \, \frac{\Delta^2(\Lambda)}{\Delta(\Lambda^2)} \, e^{-N\tr(-g_2\Lambda^2 + g_4 \Lambda^4)} \cdot
     \det \mathring{e}^{\,Ng_r\ket{R}\bra{\hspace{1pt}\Lambda^{\smash{2}}\hspace{1pt}}} \, .
     \label{partition function}
\end{align}
Repeating the argument from \cite{Kanomata:2022pdo}, and having in mind that $R$ is taken in \eqref{curvature} to be
\begin{equation}
    R = -\frac{16}{N} \diag\lr{1,2,\ldots N},
\end{equation}
we can rewrite \eqref{partition function} as
\begin{equation}
    Z =
    \int [d\Lambda] \, \mathcal{A}(\Lambda) \cdot\sum_{\mathclap{\pi \in S_N}} \sgn\pi \prod_{i=1}^N \exp(-16 g_r i \lambda_{\pi(i)}^2) \, .
\end{equation}
The function $\mathcal{A}(\Lambda)$, 
\begin{equation}
    \mathcal{A}(\Lambda) = \frac{1}{(16g_r)^{N(N-1)/2}}
    \cdot
    \frac{\Delta^2(\Lambda)}{\Delta(\Lambda^2)}
    \cdot
    e^{-N\tr(-g_2\Lambda^2 + g_4 \Lambda^4)} \, ,
\end{equation}
is a asymmetric in eigenvalues $\lambda_i$ due to the $\Delta(\Lambda^2)$-factor.
If we now perform a change of variables
\begin{equation}
    \lambda_{\pi(i)} \to \lambda_i \, ,
\end{equation}
whose Jacobian is equal to 1, we get
\begin{align}
    Z 
    &=
    \int [d\Lambda] \, (\sgn\pi\cdot \mathcal{A}(\Lambda)) \sum_{\mathclap{\pi \in S_N}} \sgn\pi \prod_{i=1}^N \exp(-16 g_r i \lambda_i^2)
    \\
    &=
    \int [d\Lambda] \, \mathcal{A}(\Lambda) \sum_{\mathclap{\pi \in S_N}} \sgn^2\pi \cdot \exp(-\sum_i 16 g_r i \lambda_i^2)
    \\
    &=
    N! \int [d\Lambda] \, \mathcal{A}(\Lambda) \; e^{\,N\tr g_r R\Lambda^2}
    \\
    &=
    \frac{N!}{(16g_r)^{N(N-1)/2}} \int [d\Lambda] \, \frac{\Delta^2(\Lambda)}{\Delta(\Lambda^2)} \, e^{-N\tr(-g_2\Lambda^2 - g_r R\Lambda^2 + g_4 \Lambda^4)} \, .
    \label{delta-problem}
\end{align}
We can now absorb the Vandermonde determinants $\Delta$ into the exponent to obtain the effective action
\begin{equation}
    S(\Lambda) = N\tr(-g_2\Lambda^2 - g_r R\Lambda^2 + g_4 \Lambda^4) - \log\frac{\Delta^2(\Lambda)}{\Delta(\Lambda^2)} + \text{const.}
    \label{log-problem}
\end{equation}
in which we have, essentially, replaced $\tr(g_r R\Phi^2)$ by $\tr(g_r R\Lambda^2)$.
Using the saddle point method and 
\begin{equation}
    \frac{\partial}{\partial\lambda_i} \log\frac{\Delta^2(\Lambda)}{\Delta(\Lambda^2)} 
    = 
    \sum_{\mathclap{\substack{j,k \\ \;\, \scriptscriptstyle 1 \leq j<k \leq N}}} 
    \frac{\partial}{\partial\lambda_i}  \log\frac{\lambda_k - \lambda_j}{\lambda_k + \lambda_j}
    =
    \sum_j \frac{2\lambda_j}{\lambda_i^2 - \lambda_j^2} \, ,
    \label{log-delta}
\end{equation}
we get
\begin{equation}
    -g_2\lambda_i + g_r \frac{i}{N} \lambda_i + 2g_4\lambda_i^3 
    =
    \frac{1}{N} \sum_j \frac{\lambda_j}{\lambda_i^2 - \lambda_j^2} \, ,
\end{equation}
that is, in the continuous limit,
\begin{equation}
    -g_2\lambda + g_r n(\lambda)\lambda + 2g_4\lambda^3 = \int d\lambda' \frac{\lambda'\rho(\lambda')}{\lambda^2-\lambda'^2} \, ,
    \label{e-distr}
\end{equation}
where 
\begin{equation}
    \frac{i}{N} \to n(\lambda) \, .
\end{equation}
Possible choices for the eigenvalue counter $n(\lambda)$ are
\begin{equation}
    n(\lambda) = \int\limits_{-r}^\lambda d\lambda' \rho(\lambda')
    \qquad
    \text{and}
    \qquad
    n(\lambda) = \frac{\lambda - (-r)}{2r} = \frac{\lambda + r}{2r} \, ,
\end{equation}
where $r$ is the radius of the eigenvalue-distribution support; the first choice includes eigenvalue multiplicities while the second does not. 

Now, the numerical simulations, which use the starting form \eqref{matrix model} of the action, imply the existence of the 1-cut-sym phase, for which the r.h.s. of \eqref{e-distr} disappears (integral of an odd function over an even support), indicating that we are doing something wrong in our derivation. After retracing our steps more carefully, we see the problem is that $\Delta(\Lambda^2)$ is not positive for all choices of $\Lambda$, precluding its absorption in the exponent in \eqref{delta-problem} since this would lead to the negative $\log$-argument and breakdown of \eqref{log-problem}. 

This is why we had to resort to the HCIZ-expansion in powers of $g_r$ to obtain the model's eigenvalue distribution. 

\section{HCIZ expansion proof}
\label{section:proof}

\setlength\arraycolsep{5pt}
\def\arraystretch{1.5}

Let us, as an example, prove \eqref{trAtrB}. Each column $\col_i$ of the matrix
\begin{equation}
\det \mathring{e}^{\,t\ket{a}\bra{\hspace{1pt}b\hspace{1pt}}} = 
\begin{pmatrix} 
    e^{ta_1b_1} & e^{ta_1b_2} & \cdots & e^{ta_1b_N} 
    \\
    e^{ta_2b_1} & e^{ta_2b_2} & \cdots & e^{ta_2b_N} 
    \\
    \vdots & \vdots & & \vdots 
    \\
    e^{ta_Nb_1} & e^{ta_Nb_2} & \cdots & e^{ta_Nb_N} 
\end{pmatrix}
\end{equation}
can be expanded in $t$ as
\begin{align}
    \col_i 
    &=  
    \begin{pmatrix} 1 \\ 1 \\ \vdots \\ 1   \end{pmatrix}
    + t \begin{pmatrix} a_1 b_i \\ a_2 b_i \\ \vdots \\  a_N b_i  \end{pmatrix}
    + \frac{t^2}{2!} \begin{pmatrix} a_1^2 b_i^2 \\ a_2^2 b_i^2 \\ \vdots \\  a_N^2 b_i^2  \end{pmatrix}
    + \cdots
    \nonumber
    \\ 
    \label{col:expansion}
    \\
    &=  
    \begin{pmatrix} 1 \\ 1 \\ \vdots \\ 1   \end{pmatrix}
    + tb_i \begin{pmatrix} a_1 \\ a_2 \\ \vdots \\  a_N  \end{pmatrix}
    + \frac{t^2b_i^2}{2!} \begin{pmatrix} a_1^2 \\ a_2^2 \\ \vdots \\  a_N^2  \end{pmatrix}
    + \cdots \, .
    \nonumber
\end{align}
Having in mind that we can split determinant along a column
\begin{equation}
    \det (\ldots, \alpha \col' + \beta \col'', \ldots) = 
    \alpha \det (\ldots,\col',\ldots) 
    +
    \beta \det (\ldots,\col'',\ldots) \, ,
\end{equation}
we can write $\det \mathring{e}^{\,t\ket{a}\bra{\hspace{1pt}b\hspace{1pt}}}$ as
\begin{align}
\det \mathring{e}^{\,t\ket{a}\bra{\hspace{1pt}b\hspace{1pt}}} 
&= \sum_{i \geq 0} t^i D_i
\nonumber
\\ 
&= \sum_{k_i \geq 0} t^{k_1+k_2+\cdots+k_N}
\frac{b_1^{k_1}b_2^{k_2} \ldots b_N^{k_N}}{k_1!k_2!\ldots k_N!}
\begin{vmatrix} 
    a_1^{k_1} & a_1^{k_2} & \cdots & a_1^{k_N} 
    \\
    a_2^{k_1} & a_2^{k_2} & \cdots & a_2^{k_N} 
    \\
    \vdots & \vdots & & \vdots 
    \\
    a_N^{k_1} & a_N^{k_2} & \cdots & a_N^{k_N} 
\end{vmatrix},
\label{det expansion}
\end{align}
where the determinant in the sum contains different choices of columns from the expansion \eqref{col:expansion}.
Notice that $k_i \neq k_j$, otherwise columns $\col_i$ and $\col_j$ are equal and the corresponding determinant vanishes. Due to this, $\det \mathring{e}^{\,t\ket{a}\bra{\hspace{1pt}b\hspace{1pt}}}$ expansion starts with $k_i$ that are permutations of $(0,1,\ldots,N-1)$,
\begin{equation}
    k_i = \pi(i) - 1 \, ,
\end{equation}
that is with
\begin{equation}
O\lr{t^{0+1+\ldots+(N-1)}} = O\lr{t^{N(N-1)/2}}.
\end{equation}
We get the second non-zero term by increasing one of the $k_i$ by 1, but still demand all $k_i$ to be different. This is satisfied only by the replacement 
\begin{equation}
    N-1 \to N \, ,    
\end{equation}
meaning that $k_i$ are permutations of $(0,1,2,\ldots,N-2,N)$. This choice yields
\begin{equation}
D_{N(N-1)/2+1} = \sum_{\pi}
\frac{\displaystyle\prod\limits_{i=1}^N b_i^{\pi(i)-1+\delta_{\pi(i),N}}}{0!1!\ldots (N-2)!N!}
\begin{vmatrix} 
    a_1^{\pi(1)-1+\delta_{\pi(i),N}} &  \cdots & a_1^{\pi(N)-1+\delta_{\pi(i),N}}
    \\
    a_2^{\pi(1)-1+\delta_{\pi(i),N}} &  \cdots & a_2^{\pi(N)-1+\delta_{\pi(i),N}}
    \\
    \vdots &  & \vdots
    \\
    a_N^{\pi(1)-1+\delta_{\pi(i),N}} &  \cdots & a_N^{\pi(N)-1+\delta_{\pi(i),N}}
\end{vmatrix}.
\end{equation}
If we order the columns by ascending powers of $a_i$, we get a factor $\sgn \pi$ from column permutations
\begin{equation}
D_{N(N-1)/2+1} =
\frac{1}{Nc_N}\sum_{\pi} \sgn\pi \prod_{i=1}^N b_i^{\pi(i)-1+\delta_{\pi(i),N}}
\begin{vmatrix} 
    a_1^{0\rlap{\phantom{N}}} & a_1^{1\phantom{N}} & \cdots & a_1^{N-2} & a_1^{N}  
    \\
    a_2^{0\rlap{\phantom{N}}} & a_2^{1\phantom{N}} & \cdots & a_2^{N-2} & a_2^{N}
    \\
    \vdots & \vdots & & \vdots &\vdots
    \\
    a_N^{0\rlap{\phantom{N}}} & a_N^{1\rlap{\phantom{N}}} & \cdots & a_N^{N-2} & a_N^{N}
\end{vmatrix}.
\end{equation}
Finally, by writing
\begin{equation}
    \mathcal{B}_{ij} = b_i^{j-1+\delta_{j,N}},
\end{equation}
we recognize
\begin{equation}
    \sum_{\pi} \sgn\pi \prod_{i=1}^N b_i^{\pi(i)-1+\delta_{\pi(i),N}} 
    =
    \sum_{\pi} \sgn\pi \prod_{i=1}^N \mathcal{B}_{i,\pi(i)}
    =
    \det \mathcal{B} \, ,
\end{equation}
that is
\begin{equation}
D_{N(N-1)/2+1} = \frac{1}{Nc_N}
\begin{vmatrix} 
    a_1^{0\rlap{\phantom{N}}} & a_1^{1\rlap{\phantom{N}}} & \cdots & a_1^{N-2} & a_1^{N}  
    \\
    a_2^{0\rlap{\phantom{N}}} & a_2^{1\rlap{\phantom{N}}} & \cdots & a_2^{N-2} & a_2^{N}
    \\
    \vdots & \vdots & & \vdots &\vdots
    \\
    a_N^{0\rlap{\phantom{N}}} & a_N^{1\rlap{\phantom{N}}} & \cdots & a_N^{N-2} & a_N^{N}
\end{vmatrix}
\cdot
\begin{vmatrix} 
    b_1^{0\rlap{\phantom{N}}} & b_1^{1\rlap{\phantom{N}}} & \cdots & b_1^{N-2} & b_1^{N}  
    \\
    b_2^{0\rlap{\phantom{N}}} & b_2^{1\rlap{\phantom{N}}} & \cdots & b_2^{N-2} & b_2^{N}
    \\
    \vdots & \vdots & & \vdots &\vdots
    \\
    b_N^{0\rlap{\phantom{N}}} & b_N^{1\rlap{\phantom{N}}} & \cdots & b_N^{N-2} & b_N^{N}
\end{vmatrix}.
\end{equation}
Determinants in the previous formula are modified Vandermonde determinants, which we denote
\begin{equation}
    \Delta(A_N|N-1 \to N)
    \quad
    \text{and}
    \quad
    \Delta(B_N|N-1 \to N) \, ,
\end{equation}
where optional subscript $N$ indicates the size of the matrix, and conditions stated after the vertical line represent the increased powers of matrix elements compared to $(0,1,\ldots,N-1)$.


Looking at the expressions for small $N$, and having in mind that $\Delta(A)|D_i$, we hypothesize that
\begin{equation}
    \Delta(A_N|N-1 \to N) =
    \begin{vmatrix} 
    1 & a_1^{\rlap{\phantom{N}}} & a_1^{2\rlap{\phantom{N}}} & \cdots & a_1^{N-2} & a_1^{N}  
    \\
    1 & a_2^{\rlap{\phantom{N}}} & a_2^{2\rlap{\phantom{N}}} & \cdots & a_2^{N-2} & a_2^{N}
    \\
    \vdots & \vdots & \vdots & & \vdots &\vdots
    \\
    1 & a_N^{\rlap{\phantom{N}}} & a_N^{2\rlap{\phantom{N}}} & \cdots & a_N^{N-2} & a_N^{N}
    \end{vmatrix}
    =
    \Delta(A_N)\tr A_N \, , 
    \label{induction:hypothesis}
\end{equation}
and then proceed to prove \eqref{induction:hypothesis} by induction. It is easy to check that \eqref{induction:hypothesis} holds for $N=2$. Let us further assume that it is true for $N$ and look what happens with the $(N+1) \times (N+1)$ matrix. We can first use the last row to eliminate the ones from the first column
\vspace{-5pt}
\begin{align}
    \Delta(A_{N+1}|N \to N+1) &=
    \begin{gmatrix}[v] 
    1 & a_1^{\rlap{\phantom{N}}} & a_1^{2\rlap{\phantom{N}}} & \cdots & a_1^{N-1} & a_1^{N+1}  
    \\ \\
    1 & a_2^{\rlap{\phantom{N}}} & a_2^{2\rlap{\phantom{N}}} & \cdots & a_2^{N-1} & a_2^{N+1}
    \\ \\
    \vdots & \vdots & \vdots & & \vdots &\vdots
    \\ \\
    1 & a_N^{\rlap{\phantom{N}}} & a_N^{2\rlap{\phantom{N}}} & \cdots & a_N^{N-1} & a_N^{N+1}
    \\ \\
    1 & a_{N+1}^{\rlap{\phantom{N}}} & a_{N+1}^{2\rlap{\phantom{N}}} & \cdots & a_{N+1}^{N-1} & a_{N+1}^{N+1}
    \rowops
    \add[\cdot\,(-1)]{8}{6}
    \add[\cdot\,(-1)]{8}{2}
    \add[\cdot\,(-1)]{8}{0}
    \end{gmatrix}
    \nonumber \\
    \\
    \nonumber \\
    &=
    \begin{gmatrix}[v] 
    0 & a_1^{\rlap{\phantom{N}}}-a_{N+1}^{\rlap{\phantom{N}}} & a_1^{2\rlap{\phantom{N}}}-a_{N+1}^{2\rlap{\phantom{N}}} & \cdots & a_1^{N-1}-a_{N+1}^{N-1} & a_1^{N+1}-a_{N+1}^{N+1} 
    \\ \\
    0 & a_2^{\rlap{\phantom{N}}}- a_{N+1}^{\rlap{\phantom{N}}} & a_2^{2\rlap{\phantom{N}}}-a_{N+1}^{2\rlap{\phantom{N}}} & \cdots & a_2^{N-1}-a_{N+1}^{N-1} & a_2^{N+1}-a_{N+1}^{N+1}
    \\ \\
    \vdots & \vdots & \vdots & & \vdots &\vdots
    \\ \\
    0 & a_N^{\rlap{\phantom{N}}}-a_{N+1}^{\rlap{\phantom{N}}} & a_N^{2\rlap{\phantom{N}}}-a_{N+1}^{2\rlap{\phantom{N}}} & \cdots & a_N^{N-1}-a_{N+1}^{N-1} & a_N^{N+1}-a_{N+1}^{N+1}
    \\ \\
    1 & a_{N+1}^{\rlap{\phantom{N}}} & a_{N+1}^{2\rlap{\phantom{N}}} & \cdots & a_{N+1}^{N-1} & a_{N+1}^{N+1}
    \end{gmatrix}\hspace{-3pt}.
    \nonumber
\end{align}


The next step is to use the Laplace expansion along the first column, followed by the subtraction of the $a_{N+1}\col_i$ from $\col_{i+1}$, except for the last column $\col_{N+1}$ from which we subtract $a_{N+1}^2\col_N^{\phantom{2}}$:
\begin{multline}
    \Delta(A_{N+1}|N \to N+1) = (-1)^{N+2} \times
    \\ 
    \times
    \begin{gmatrix}[v] 
    a_1^{\rlap{\phantom{N}}}-a_{N+1}^{\rlap{\phantom{N}}} & a_1^{2\rlap{\phantom{N}}}-a_{N+1}^{2\rlap{\phantom{N}}} & \cdots & a_1^{N-2}-a_{N+1}^{N-2} & a_1^{N-1}-a_{N+1}^{N-1} & a_1^{N+1}-a_{N+1}^{N+1} 
    \\ \\
    a_2^{\rlap{\phantom{N}}}-a_{N+1}^{\rlap{\phantom{N}}} & a_2^{2\rlap{\phantom{N}}}-a_{N+1}^{2\rlap{\phantom{N}}} & \cdots & a_2^{N-2}-a_{N+1}^{N-2} & a_2^{N-1}-a_{N+1}^{N-1} & a_2^{N+1}-a_{N+1}^{N+1}
    \\ \\
    \vdots & \vdots & & \vdots & \vdots & \vdots
    \\ \\
    a_N^{\rlap{\phantom{N}}}-a_{N+1}^{\rlap{\phantom{N}}} & a_N^{2\rlap{\phantom{N}}}-a_{N+1}^{2\rlap{\phantom{N}}} & \cdots & a_N^{N-2}-a_{N+1}^{N-2} & a_N^{N-1}-a_{N+1}^{N-1} & a_N^{N+1}-a_{N+1}^{N+1}
    \colops
    \add[\cdot\,(-a_{N+1})]{0}{1}
    \add[\cdot\,(-a_{N+1})]{3}{4}
    \add[\cdot\,(-a_{N+1}^2)]{4}{5}
    \end{gmatrix}\hspace{-3pt}.
\end{multline}
We can simplify the $(i,j+1)$ element of the determinant into
\begin{equation}
    \lr{a_i^{j+1} - a_{N+1}^{j+1}} - a_{N+1}^{\rlap{\phantom{j}}}\lr{a_i^j - a_{N+1}^j} = 
    a_i^j\lr{a_i^{\phantom{j}} - a_{N+1}^{\phantom{j}}},
\end{equation}
and $(i,N+1)$ element into
\begin{equation}
    \lr{a_i^{N+1} - a_{N+1}^{N+1}} - a_{N+1}^{2\rlap{\phantom{N}}}\lr{a_i^{N-1} - a_{N+1}^{N-1}} = 
    a_i^{N-1}\lr{a_i^{2\vphantom{N}} - a_{N+1}^{2{\vphantom{N}}}},
\end{equation}
which gives
\begin{multline}
    \Delta(A_{N+1}|N \to N+1) = (-1)^{N+2} \times
    \\ \\ 
    \times
    \begin{gmatrix}[v] 
    a_1^{\rlap{\phantom{N}}}-a_{N+1}^{\rlap{\phantom{N}}} & a_1^{\rlap{\phantom{N}}}\!\left(a_1^{\llap{\phantom{N}}} - a_{N+1}^{\rlap{\phantom{N}}}\right) & \cdots & 
    a_1^{N-2\rlap{\phantom{N}}}\!\left(a_1^{\rlap{\phantom{N}}} - a_{N+1}^{\rlap{\phantom{N}}}\right) & \,\,a_1^{N-1}\!\left(a_1^{2\rlap{\phantom{N}}} - a_{N+1}^{2\rlap{\phantom{N}}}\right) 
    \\ \\
    a_2^{\rlap{\phantom{N}}}-a_{N+1}^{\rlap{\phantom{N}}} & a_2^{\rlap{\phantom{N}}}\!\left(a_2^{\rlap{\phantom{N}}} - a_{N+1}^{\rlap{\phantom{N}}}\right) & \cdots & 
    a_2^{N-2\rlap{\phantom{N}}}\!\left(a_2^{\rlap{\phantom{N}}} - a_{N+1}^{\rlap{\phantom{N}}}\right) & \,\,a_2^{N-1}\!\left(a_2^{2\rlap{\phantom{N}}} - a_{N+1}^{2\rlap{\phantom{N}}}\right)
    \\ \\
    \vdots & \vdots & & \vdots & \vdots
    \\ \\
    a_N^{\rlap{\phantom{N}}}-a_{N+1}^{\rlap{\phantom{N}}} & a_N^{\rlap{\phantom{N}}}\!\left(a_N^{\rlap{\phantom{N}}} - a_{N+1}^{\rlap{\phantom{N}}}\right) & \cdots & 
    a_N^{N-2\rlap{\phantom{N}}}\!\left(a_N^{\rlap{\phantom{N}}} - a_{N+1}^{\rlap{\phantom{N}}}\right) & \,\,a_N^{N-1}\!\left(a_N^{2\rlap{\phantom{N}}} - a_{N+1}^{2\rlap{\phantom{N}}}\right)
    \end{gmatrix}\hspace{-7pt}.\hspace{-5pt}
\end{multline}
The each element in the $i^\text{th}$ row has a common multiplier $a_i^{\phantom{j}} - a_{N+1}^{\phantom{j}}$, which we can extract from the determinant to get
\begin{multline}
    \Delta(A_{N+1}|N \to N+1) = (-1)^{N+2} \prod_{j=1}^N \left(a_i^{\phantom{j}} - a_{N+1}^{\phantom{j}}\right) \times 
    \\ 
    \times
    \begin{gmatrix}[v] 
    1 & a_1^{\rlap{\phantom{N}}} & \cdots & 
    a_1^{N-2\rlap{\phantom{N}}} & \,\,a_1^{N-1}\!\left(a_1^{\rlap{\phantom{N}}} + a_{N+1}^{\rlap{\phantom{N}}}\right) 
    \\ \\
    1 & a_2^{\rlap{\phantom{N}}} & \cdots & 
    a_2^{N-2\rlap{\phantom{N}}} & \,\,a_2^{N-1}\!\left(a_2^{\rlap{\phantom{N}}} + a_{N+1}^{\rlap{\phantom{N}}}\right)
    \\ \\
    \vdots & \vdots & & \vdots & \vdots
    \\ \\
    1 & a_N^{\rlap{\phantom{N}}} & \cdots & 
    a_N^{N-2\rlap{\phantom{N}}} & \,\,a_N^{N-1}\!\left(a_N^{\rlap{\phantom{N}}} + a_{N+1}^{\rlap{\phantom{N}}}\right)
    \end{gmatrix}\hspace{-3pt}.
\end{multline}
We can now split the determinant along the last column 
\begin{equation}
    \begin{gmatrix}[v] 
    1 & a_1^{\rlap{\phantom{N}}} & \cdots & 
    a_1^{N-2\rlap{\phantom{N}}} & a_1^{N\phantom{1\!\!\!}}
    \\ \\
    1 & a_2^{\rlap{\phantom{N}}} & \cdots & 
    a_2^{N-2\rlap{\phantom{N}}} & a_2^{N\phantom{1\!\!\!}}
    \\ \\
    \vdots & \vdots & & \vdots & \vdots
    \\ \\
    1 & a_N^{\rlap{\phantom{N}}} & \cdots & 
    a_N^{N-2\rlap{\phantom{N}}} & a_N^{N\phantom{1\!\!\!}}
    \end{gmatrix}
    +
    \hspace{7pt}
    a_{N+1}^{\rlap{\phantom{N}}}
    \begin{gmatrix}[v] 
    1 & a_1^{\rlap{\phantom{N}}} & \cdots & 
    a_1^{N-2\rlap{\phantom{N}}} & a_1^{N-1} 
    \\ \\
    1 & a_2^{\rlap{\phantom{N}}} & \cdots & 
    a_2^{N-2\rlap{\phantom{N}}} & a_2^{N-1}
    \\ \\
    \vdots & \vdots & & \vdots & \vdots
    \\ \\
    1 & a_N^{\rlap{\phantom{N}}} & \cdots & 
    a_N^{N-2\rlap{\phantom{N}}} & a_N^{N-1}
    \end{gmatrix}\hspace{-3pt},
\end{equation}
and end  the proof by writing 
\begin{align}
    \Delta(A_{N+1}|N \to N+1) 
    &=
    \prod_{j=1}^N \left(a_{N+1}-a_i\right) 
    \left(\Delta(A_N|N-1 \to N) + a_{N+1}\Delta(A_N)\right)
    \label{recursion}
    \\
    &=
    \prod_{j=1}^N \left(a_{N+1}-a_i\right) 
    \left(\Delta(A_N)\tr A_N + a_{N+1}\Delta(A_N)\right)
    \\
    &=
    \prod_{j=1}^N \left(a_{N+1}-a_i\right)\Delta(A_N) 
    \left(\tr A_N + a_{N+1}\right)
    \\
    &=
    \Delta(A_{N+1}) \tr A_{N+1} \, ,
\end{align}
where in the second line we used our induction hypothesis \eqref{induction:hypothesis}.

This proof can be straightforwardly modified for other terms in the expansion of HCIZ. The key step is to identify the non-vanishing combinations of $k_i$ in \eqref{det expansion} once we increase the power of $t$ by $p$. For the first few terms, it is easy to identify a small number of the possible partitions of $p$ into a sum of positive numbers $p_i$, and then try to increase the highest values in $(0,1,\ldots,N-1)$ by $p_i$ in such a way to keep all $k_i$ different. For $S_2$ we need permutations of $(0,1,\ldots,N-2,N+1)$ and $(0,1,\ldots,N-3,N-1,N)$, and for $S_3$ the permutations of $(0,1,\ldots,N-2,N+2)$, $(0,1,\ldots,N-3,N-1,N+1)$, and $(0,1,\ldots,N-4,N-2,N-1,N)$.

\section{Fourth order eigenvalue-distribution equation}
\label{appendix:4th order distribution}

Applying the saddle-point method to the obtained $O(g_r^4)$ effective action 
\begin{multline}\label{model:appD}
    S(\Lambda) = N\tr\left(
    - \left(g_2 - 8g_r \right)\Lambda^2 
    + \left(g_4 - \frac{32}{3}g_r^2\right)\Lambda^4 
    + \frac{1024}{45}g_r^4\tr\Lambda^8 
    \right)+
    \\
    + \frac{32}{3}g_r^2\tr^{\mathrlap{2}}\, \Lambda^2
    + \frac{1024}{15}g_r^4\tr^{\mathrlap{2}}\,\Lambda^4
    - \frac{4096}{45}g_r^4\tr\Lambda^6\tr\Lambda^2
    - \log\Delta^2(\Lambda) \, ,
\end{multline}
results in the following eigenvalue-distribution equation
\begin{multline}
    -\lr{g_2 - 8g_r - \frac{64}{3}g_r^2m_2 + \frac{4096}{45}g_r^4 m_6}\lambda 
    +2\lr{g_4 - \frac{32}{3}g_r^2 + \frac{2048}{15}g_r^4 m_4}\lambda^3
    -
    \\
    -\frac{4096}{15}g_r^4 m_2\lambda^5
    +\frac{4096}{45}g_r^4 \lambda^7
    =
    \underset{\mathclap{\text{support}}}{\int} d\lambda' \, \frac{\rho(\lambda')}{\lambda-\lambda'} \, .
    \label{eigen1}
\end{multline}
Using the following ansatz for the 1-cut-sym distribution
\begin{equation}
    \rho(\lambda) = \frac{1}{\pi}(\rho_0 + \rho_2\lambda^2 + \rho_4\lambda^4 + \rho_6\lambda^6)\sqrt{r^2 - \lambda^2} \, ,
    \label{eigen2}
\end{equation}
we get
\begin{multline}
    \int\limits_{-r}^{+r}  d \lambda' \, \frac{\rho(\lambda')}{\lambda-\lambda'} 
    =
    \frac{1}{16}\lr{16\rho_0 - 8\rho_2r^2 - 2\rho_4r^4 - \rho_6r^6}\lambda 
    +
    \\
    +
    \frac{1}{8}\lr{8\rho_2 - 4\rho_4r^2 - \rho_6r^4}\lambda^3
    +
    \frac{1}{2}\lr{2\rho_4 - \rho_6r^2}\lambda^5
    +\rho_6\lambda^7,
    \label{eigen3}
\end{multline}
and
\begin{equation}
    \int\limits_{-r}^{+r}  d \lambda' \, \rho(\lambda') 
    = 
    \frac{r^2}{128}\lr{64\rho_0 + 16\rho_2r^2 + 8\rho_4r^4 + 5\rho_6r^6}
    =
    1 \, .
    \label{eigen4}
\end{equation}
Equating coefficients in \eqref{eigen1} and \eqref{eigen3}, leads to 
\begin{subequations}
\label{rho_i}
\begin{multline}
    \rho_0 = (g_4r^2 - g_2) + 8g_r + \frac{32}{3}g_r^2(2m_2 - r^2) +
    \\
    + \frac{256}{45}g_r^4(5r^6 - 18m_2r^4 + 24m_4r^2 - 16m_6) \, ,
\end{multline}
\begin{equation}
    \rho_2 = 2g_4 - \frac{64}{3}g_r^2 + \frac{512}{15}g_r^4(r^4 - 4m_2r^2 + 8m_4) \, ,
\end{equation}
\begin{equation}
    \rho_4 = \frac{2048}{45}g_r^4(r^2 - 6m_2) \, ,
    \qquad\qquad\!\!\!
    \rho_6 = \frac{4096}{45}g_r^4 \, ,
\end{equation}
\end{subequations}
where the normalized moments $m_k$
\begin{equation}
    m_k = \int\limits_{-r}^{+r}  d\lambda' \, \rho(\lambda')\, \lambda'^{\,k} ,
\end{equation}
satisfy
\begin{subequations}
\label{m_i}
\begin{equation}
    m_2 = \frac{r^4}{\,\,256\,\,}(32\rho_0 + 16\rho_2r^2 + 10\rho_4r^4 + \phantom{2}7\rho_6r^6) \, ,
\end{equation}
\begin{equation}
    m_4 = \frac{r^6}{1024}(64\rho_0 + 40\rho_2r^2 + 28\rho_4r^4 + 21\rho_6r^6) \, ,
\end{equation}
\begin{equation}
    m_6 = \frac{r^8}{2048}(80\rho_0 + 56\rho_2r^2 + 42\rho_4r^4 + 33\rho_6r^6) \, .
\end{equation}
\end{subequations}
A seven-equation system (\ref{rho_i},\,\ref{m_i}) can be solved in $\rho_i$ and $m_i$, resulting in unilluminating expressions dependent on the action parameters and the distribution radius. After their substitution, the phase transition condition $\rho_0 = 0$ and the distribution normalization condition respectively become
\begin{multline}
    (g_4r^2 - g_2) 
    + 8g_r 
    + \frac{8}{3}g_r^2r^2(2g_4r^4 - g_2r^2 -4)
    +
    \\
    + \frac{64}{3}g_r^3r^4
    - \frac{32}{45}g_r^4r^6(g_4r^4 - g_2r^2 +40)
    = 0 \, ,
    \label{transition condition}
\end{multline}
\begin{multline}
    \frac{1}{4}r^2(3g_4r^2 - 2g_2) 
    + 4g_r r^2
    + \frac{4}{3}g_r^2r^4(2g_4r^4 - g_2r^2 -6)
    +
    \\
    + \frac{32}{3}g_r^3r^6
    - \frac{8}{45}g_r^4r^8(23g_4r^4 - 14g_2r^2 +20)
    = 1 \, ,
    \label{normalization condition}
\end{multline}
up to $O(g_r^4)$. By writing
\begin{equation}
    r^2 = r_0^2 + \sum_{i=1}^\infty g_r^ir_i^2 \, ,  
\end{equation}
we can now solve \eqref{normalization condition} order-by-order in $g_r$ to get
\begin{equation}
    r_0^2 = \frac{g_2 + \sqrt{g_2^2 + 12g_4}}{3g_4} \, ,
    \qquad\quad
    r_1^2 = -\frac{8}{3g_4}\left(1 + \frac{g_2}{\sqrt{g_2^2+12 g_4}}\right),
\end{equation}
and lengthy expressions for $r_2^2$, $r_3^2$ and $r_4^2$. As a sanity check, the expression for $r_0^2$ is exactly what we expect in the PP model.
Plugging these into the transition line condition \eqref{transition condition}, writing
\begin{equation}
    g_2 = g_2^{(0)} + \sum_{i=1}^\infty g_r^ig_2^{(i)},
\end{equation}
and solving it again order-by-order in $g_r$, we finally arrive at the following expression for $O(g_r^4)$ transition line
\begin{equation}\label{e16}
    g_2 = 2\sqrt{g_4} + 8g_r + \frac{32}{3}\frac{g_r^2}{\sqrt{g_4}} + \frac{256}{15}\frac{g_r^4}{g_4\sqrt{g_4}} \, .
\end{equation}

In a slightly different fashion, we can obtain a numerical solution to the phase transition of the model \eqref{model:appD}, i.e. to conditions similar to (\ref{transition condition},\,\ref{normalization condition}) but with all the $g_r$ terms, by choosing a numerical value of $g_4$ and solving the equations for $g_2$ and $r$. By a dense choice of the $g_4$-values, we can obtain a reasonably good approximation to the exact transition line of the model \eqref{SRPP}, which is shown by the green dotted line in Figure \ref{figure:O4}.

\section{Numerical result extrapolation}
\label{appendix:extrapolation}

We here present the details of the $N \to \infty$ extrapolation of the numerical results.

\begin{figure}[t]
\centering  
\includegraphics[width=1.00\textwidth]{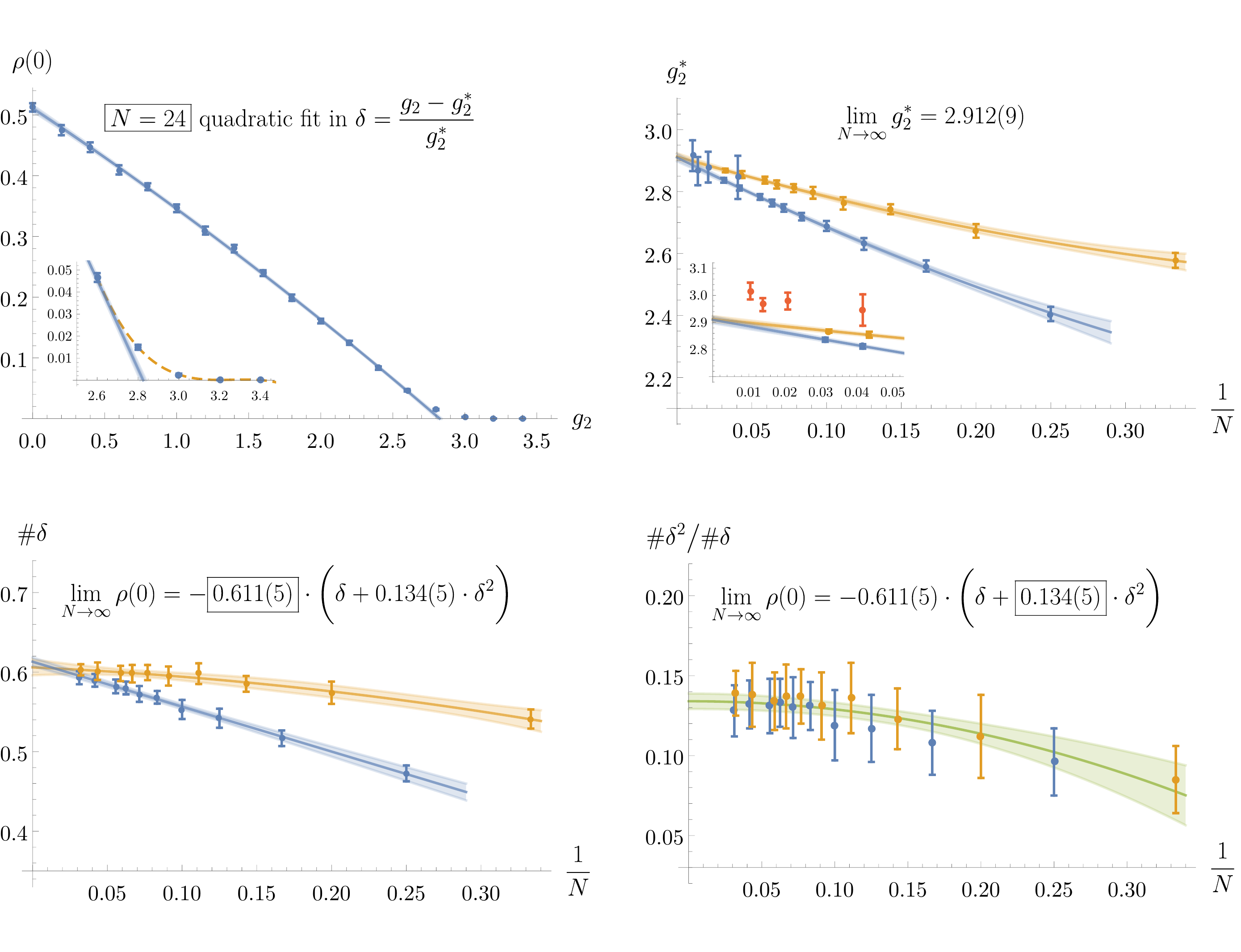}
\caption{
Extrapolation of the transition-point location $g_2^*$ in the large-$N$ limit for $(g_4,\,g_r)=(1,\,0.1)$. Blue data points have even $N$, orange data points have odd $N$, and red data points are generated by the eigenvalue-flipping algorithm. The green fit in the bottom-left plot goes through both blue and orange points since they do not seem to belong to separate groups at this resolution.
}
\label{figure:extrapolation}
\end{figure}

As already stated by \eqref{rho0order2}, the mid-point of the eigenvalue distribution should decrease nearly linearly with $\delta$ as we approach the phase transition
\begin{equation}
    \rho(0) = -\frac{\sqrt[4]{4g_4}}{\pi}
    \lr{1 + \frac{4g_r}{\sqrt{g_4}} - \frac{8g_r^2}{3g_4}}
    \cdot
    \lr{
    \delta
    +
    \frac{1}{8}\lr{1 + \frac{4g_r}{\sqrt{g_4}} - \frac{128g_r^2}{3g_4}}\delta^2
    }.
\end{equation}
This is nicely illustrated for $N=24$ in the top-left plot in Figure \ref{figure:extrapolation}. The tail in the zoomed-in region is due to finite-$N$ effects. The fit is performed by including more and more points until the tail becomes apparent in the plot of normalized fit-residuals.

We then collect the transition point position $g_2^*$, the linear coefficient $\#\delta$
\begin{equation}
    \#\delta 
    \;\leftrightarrow\; 
    \frac{\sqrt[4]{4g_4}}{\pi}
    \lr{1 + \frac{4g_r}{\sqrt{g_4}} - \frac{8g_r^2}{3g_4}},
\end{equation}
and the ratio of the quadratic and the linear coefficient $\#\delta^2/\#\delta$
\begin{equation}
    \frac{\,\#\delta^2}{\,\#\delta^{\phantom{2}}} 
    \;\leftrightarrow\;  
    \frac{1}{8}\lr{1 + \frac{4g_r}{\sqrt{g_4}} - \frac{128g_r^2}{3g_4}},
\end{equation}
for various values of $N$, and perform their extrapolation for $N \to \infty$. Luckily, fits at most quadratic in $1/N$ were sufficient in this regard.

Theoretical prediction from \eqref{4th order transition line} agrees exceptionally well with the numerical value for the transition point at $(g_r,\,g_4) = (0.1, \,1)$:
\begin{subequations}
\begin{align}
\text{theoretical:} \qquad  g_2^* &\approx 2.908 \, ,
\\
\text{numerical:} \qquad  g_2^* &\approx 2.912(9) \, .
\end{align}
\end{subequations}
The coefficients of $\delta$ and $\delta^2$ terms are also close to the prediction of \eqref{rho0order2}:
\begin{subequations}
\begin{align}
\text{theoretical:} \qquad \rho(0) &\approx -0.618\cdot\lr{\delta + 0.122\cdot\delta^2},
\\
\text{numerical:} \qquad \rho(0) &\approx -0.611(5)\cdot\lr{\delta + 0.134(5)\cdot\delta^2}.
\end{align}
\end{subequations}

Blue data points in Figure \ref{figure:extrapolation} come from even $N$ and orange data points from odd $N$, which need to be treated separately due to a small local $\rho$-maximum at $\lambda=0$. The data are gathered for matrix sizes up to $N=96$.

Red points in the zoomed-in region of the upper-right plot come from the eigenvalue-flipping algorithm and have a systematic shift compared to the Hybrid Monte Carlo data fit. This is presumably due to overestimated contribution of false vacua corresponding to 2-cut-asym solutions. One of their cuts is bigger than the other, and it presumably has a wider support than in the symmetric case, similarly to \cite{Tekel:2017nzf}. Its tails could then over-contribute to the $\lambda=0$ in eigenvalue-histograms, shifting the apparent phase transition point to a larger $g_2$. It would be interesting to investigate this in more detail. Once the shift is compensated, the data are included in the fit (four blue data points with large uncertainties).

\bibliographystyle{JHEP}
\bibliography{bibliography}

\end{document}